\documentclass[letter,bibyear]{aa} 

\usepackage{epsfig}
\usepackage{amsmath}
\usepackage{subfigure}
\usepackage{natbib}
\usepackage{multicol}
\usepackage{multirow}
\usepackage{color}
\usepackage{float}
\usepackage{longtable}
\usepackage{graphicx}
\usepackage{epstopdf}
\usepackage{booktabs}

\usepackage{stackrel,amssymb,amsmath}
\newcommand{\leftrarrows}{\mathrel{\raise.75ex\hbox{\oalign{%
  $\scriptstyle\leftarrow$\cr
  \vrule width0pt height.5ex$\hfil\scriptstyle\relbar$\cr}}}}
\newcommand{\lrightarrows}{\mathrel{\raise.75ex\hbox{\oalign{%
  $\scriptstyle\relbar$\hfil\cr
  $\scriptstyle\vrule width0pt height.5ex\smash\rightarrow$\cr}}}}
\newcommand{\Rrelbar}{\mathrel{\raise.75ex\hbox{\oalign{%
  $\scriptstyle\relbar$\cr
  \vrule width0pt height.5ex$\scriptstyle\relbar$}}}}

\usepackage{txfonts}

\newcommand{\jms}{J.~Mol.~Spectrosc.}   
\newcommand{\jmst}{J.~Mol.~Struct.}   

\newcommand{\kms}{km s$^{-1}$}

\newcommand{\prev}{Phys. Rev.}
\begin{document}

\title{The magnesium paradigm in IRC+10216:\\ Discovery of MgC$_4$H$^+$, MgC$_3$N$^+$,
MgC$_6$H$^+$, and MgC$_5$N$^+$ \thanks{Based on observations carried out with the Yebes\,40m telescope 
(projects 19A010, 20A017, 20B014, 21A019) and IRAM\,30m telescope. 
The 40m radio telescope at Yebes Observatory is operated by the Spanish Geographic 
Institute (IGN, Ministerio de Fomento). IRAM is supported by INSU/CNRS (France), MPG (Germany), and IGN (Spain).}}

\author{
J.~Cernicharo\inst{1},
C.~Cabezas\inst{1},
J.~R.~Pardo\inst{1},
M.~Ag\'undez\inst{1},
O.~Roncero\inst{1},
B.~Tercero\inst{2,3},
N.~Marcelino\inst{2,3},
M.~Gu\'elin\inst{4},
Y.~Endo\inst{5}, and
P.~de~Vicente\inst{2}
}
\institute{Grupo de Astrof\'isica Molecular, Instituto de F\'isica Fundamental (IFF-CSIC),
C/ Serrano 121, 28006 Madrid, Spain\\ \email jose.cernicharo@csic.es \& carlos.cabezas@csic.es
\and Centro de Desarrollos Tecnol\'ogicos, Observatorio de Yebes (IGN), 19141 Yebes, Guadalajara, Spain
\and Observatorio Astron\'omico Nacional (OAN, IGN), Madrid, Spain
\and Institut de Radioastronomie Millim\'etrique, 300 rue de la Piscine, F-38406,
  Saint Martin d'H\`eres, France
\and Department of Applied Chemistry, Science Building II, National Yang Ming Chiao Tung University, 1001 Ta-Hsueh Rd., Hsinchu 300098, Taiwan
}

\date{Received 20/3/2023; accepted 5/4/2023}

\abstract{We found four series of harmonically related lines in IRC\,+10216 with the Yebes\,40m and IRAM\,30m telescopes.
The first series corresponds to a molecule with
a rotational constant, $B$, of 1448.5994$\pm$0.0013 MHz and a distortion constant, $D$, of 63.45$\pm$1.15 Hz
and covers upper quantum numbers from $J_u$=11 up to 33 (B1449). 
The second series is fitted with $B$=1446.9380$\pm$0.0098 MHz and $D$=91$\pm$23 Hz and covers upper quantum numbers from
$J_u$=11 up to 17 (B1447).
The third series is fitted with $B$=598.7495$\pm$0.0011 
MHz and D=6.13$\pm$0.43 Hz and covers quantum numbers from $J_u$=26 up to 41 (B599). 
Finally, the frequencies of the last series of lines can be reproduced with $B$=594.3176$\pm$0.0026 MHz and $D$=4.92$\pm$1.16 Hz (B594). 
The large values of $D$ point toward four metal-bearing carriers. After exploring
all plausible candidates containing Na, Al, Mg, and other metals, our ab initio calculations indicate that
the cations MgC$_4$H$^+$, MgC$_3$N$^+$, MgC$_6$H$^+$, and MgC$_5$N$^+$ must be the carriers of B1449, B1447, B599, and B594, respectively. These cations could be formed by the radiative association of
Mg$^+$ with C$_4$H, C$_3$N, C$_6$H, and C$_5$N, respectively. 
We calculated the radiative association rate coefficient of Mg$^+$ with C$_4$H, C$_3$N, C$_6$H, and C$_5$N
and incorporated them in our chemical model. The results confirm that the
Mg-bearing cations can be formed through these radiative association reactions 
in the outer layers of IRC\,+10216. This is the first time that cationic metal-bearing species 
have been found in space. These results provide a new paradigm on the reactivity of ionized metals 
with abundant radicals and open the door for further characterization of similar species in 
metal-rich astrophysical environments.}


\titlerunning{MgC$_4$H$^+$ and MgC$_6$H$^+$ in IRC\,+10216}
\authorrunning{Cernicharo et al.}

\maketitle

\section{Introduction}
Polyatomic metal-containing molecules are detected only toward the
circumstellar envelopes of evolved stars. Diatomic molecules with Al,
Na, or K, and a halogen atom have been detected in the prototypical carbon star envelope IRC\,+10216
\citep{Cernicharo1987}. These species are formed in the hot inner parts of the envelope,
close to the star. Metals such as Na, K, Ca, Cr, and Fe are found to survive in the gas phase as
neutral and/or ionized atoms in the outer layers of IRC\,+10216  \citep{Mauron2010}, which points
to a very rich metal chemistry in the cool outer envelope. Various triatomic metal-bearing species
containing Mg, Na, Al, K, Fe, and Ca have been detected in the external layers of IRC\,+10216
\citep{Kawaguchi1993,Turner1994,Ziurys1995,Ziurys2002,Pulliam2010,Zack2011,Cernicharo2019a,Changala2022}. 
To date, the only metal found to  lead to molecules with more than three atoms in IRC\,+10216 is 
magnesium.
These species are
MgCCH \citep{Agundez2014,Cernicharo2019b}, 
HMgNC \citep{Cabezas2013}, MgC$_4$H, MgC$_3$N \citep{Cernicharo2019b}, MgC$_6$H, MgC$_5$N \citep{Pardo2021}, HMgCCCN, and NaCCCN \citep{Cabezas2023}. 
The observed abundances for these species are well reproduced adopting the
formation mechanism proposed by \citet{Dunbar2002}. Briefly, ionized metal atoms (M$^+$) react 
with large polyynes (C$_{2n}$H$_2$) and cyanopolyynes (HC$_{2n+1}$N) to form, 
through radiative association, the cationic complexes MC$_{2n}$H$_2$$^+$ and MHC$_{2n+1}$N$^+$, 
respectively, which then recombine dissociatively with electrons yielding as fragments the 
$^2\Sigma$ radicals MC$_{2n}$H and MC$_{2n+1}$N.

In this letter we report the detection of four new molecules in the direction of IRC\,+10216. On the basis of
precise quantum chemical calculations, exploration of all possible candidates, and chemical
modeling, we assign them to the cations MgC$_4$H$^+$, MgC$_3$N$^+$, MgC$_6$H$^+$, and MgC$_5$N$^+$. 

\section{Observations} \label{observations}

New receivers, built within the Nanocosmos project\footnote{ERC grant ERC-2013-Syg-610256-NANOCOSMOS.\\
https://nanocosmos.iff.csic.es/} and installed at the Yebes 40m radiotelescope, were used
for the observations of IRC\,+10216
($\alpha_{J2000}=9^{\rm h} 47^{\rm  m} 57.36^{\rm s}$ and $\delta_{J2000}=
+13^\circ 16' 44.4''$).
These observations belong to the Nanocosmos survey of evolved stars \citep{Pardo2022}. Additional 
details on these observations are given in Appendix \ref{line_parameters}.

\section{Results} \label{results}
The latest Q-band Yebes\,40m data between 31 and 50 GHz of IRC\,+10216 have a sensitivity of 0.15-0.32 mK 
in the $T_A^*$ scale for a spectral resolution of 220 kHz, and show a forest of sub-mK unidentified lines. 
Line identification 
in this work was done using the  catalogs 
MADEX \citep{Cernicharo2012}, CDMS \citep{Muller2005}, and JPL \citep{Pickett1998}. 
By January 2023 the MADEX code contained 6464 spectral
entries corresponding to different vibrational states and isotopologs of 1765 molecules.

\subsection{Detection of B1449 (MgC$_4$H$^+$) and B1447 (MgC$_3$N$^+$)} \label{det_mgc4h+}

\begin{figure}
\centering
\includegraphics[width=0.49\textwidth,angle=0]{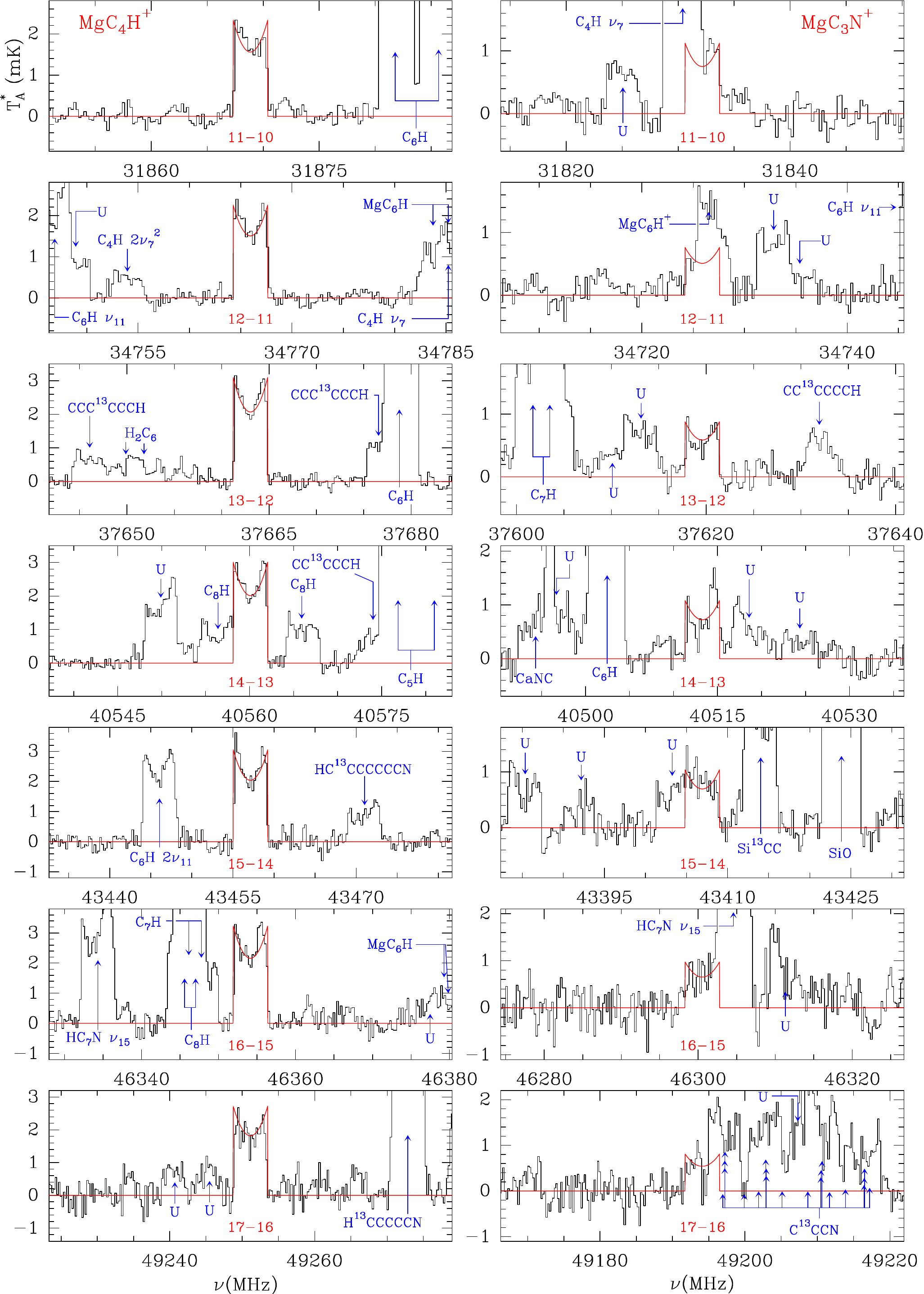}
\caption{Observations with the Yebes\,40m telescope
toward IRC\,+10216. The left and right panels show the lines of 
B1449 and B1447 (MgC$_4$H$^+$ and MgC$_3$N$^+$), respectively.
The line parameters are given in Table \ref{line_parameters_mgc4h+}
and Table \ref{line_parameters_mgc3n+}, respectively.
The abscissa corresponds to the rest frequency assuming a local standard of rest velocity of $-$26.5
km s$^{-1}$. 
The ordinate is the antenna temperature corrected for atmospheric and telescope losses in millikelvin.
The red lines show the fitted line profiles adopting a terminal velocity of expansion of
14.5 \kms \citep{Cernicharo2000}.
}
\label{fig_mgc4h+}
\end{figure}

Among the strongest IRC\,+10216 unidentified lines in the Q band, seven  appear in 
perfect harmonic relation from $J_u$=11 to $J_u$=17. They exhibit large intensities, $\sim$2-3 mK, and are shown 
in the left panels of Fig. \ref{fig_mgc4h+}.
Further exploration of the IRAM 30m radio telescope 3mm data reveals
nine additional lines, previously unidentified, following the
same harmonic relation up to $J_u$=33 (see Fig. \ref{fig_mgc4h+_3mm}). 
The derived line parameters for this series, named B1449, are provided in Table 
\ref{line_parameters_mgc4h+}.
The 16 lines can be fitted together to the standard
Hamiltonian of a linear rotor ($\nu$=2\,$B$\,$J_u$$-$4\,$D$\,$J_u$$^3$)
with $B$=1448.5994$\pm$0.0013 MHz and $D$=63.45$\pm$1.15 Hz. The standard
deviation of the fit is 200 kHz (60 kHz for the Q-band  lines alone). We checked that values of $B$ divided by 2-6
do not match the observations as many lines would be missing. Moreover, the lines do not show any evidence
of spectroscopic broadening (see Fig. \ref{fig_mgc4h+}) as they can be fitted with the standard profile for an 
expanding envelope with
14.5 \kms\ terminal velocity \citep{Cernicharo2000}. Hence, the carrier of B1449 has a ground 
electronic state $^1\Sigma$.

A second series of lines appear at frequencies very close to those of B1449. The lines 
are shown in the right panels of Fig. \ref{fig_mgc4h+} and
their line parameters are given in Table 
\ref{line_parameters_mgc3n+}. These lines can be
fitted to a linear $^1\Sigma$ molecule with $B$=1446.9380$\pm$0.0098 MHz and D=91$\pm$23 Hz (B1447). The standard deviation of the fit is
140 kHz. These results point to
two molecules with a $^1\Sigma$ electronic ground state, or with a $^2\Sigma$ electronic
ground state but with a very low value of the spin-rotation constant ($\gamma\le$0.2 MHz). A value for
$\gamma$ larger than 0.2 MHz would produce a detectable broadening 
given the high signal-to-noise ratio
lines of B1449 lines shown in Fig. \ref{fig_mgc4h+}.
We note that the tight harmonic relations in the frequencies of each series of lines, together with
the different line intensities of the two series, rule
out that they could both arise from the fine structure components from a single open shell molecule (e.g.,
a $^2\Sigma$ molecule with a large $\gamma$).  
A detailed inspection of the QUIJOTE line survey \citep{Cernicharo2021a,Cernicharo2022,Cernicharo2023}
shows that the lines of B1449 and B1447 are not detected in the cold starless core TMC-1.

\begin{figure*}
\centering
\includegraphics[width=1.0\textwidth]{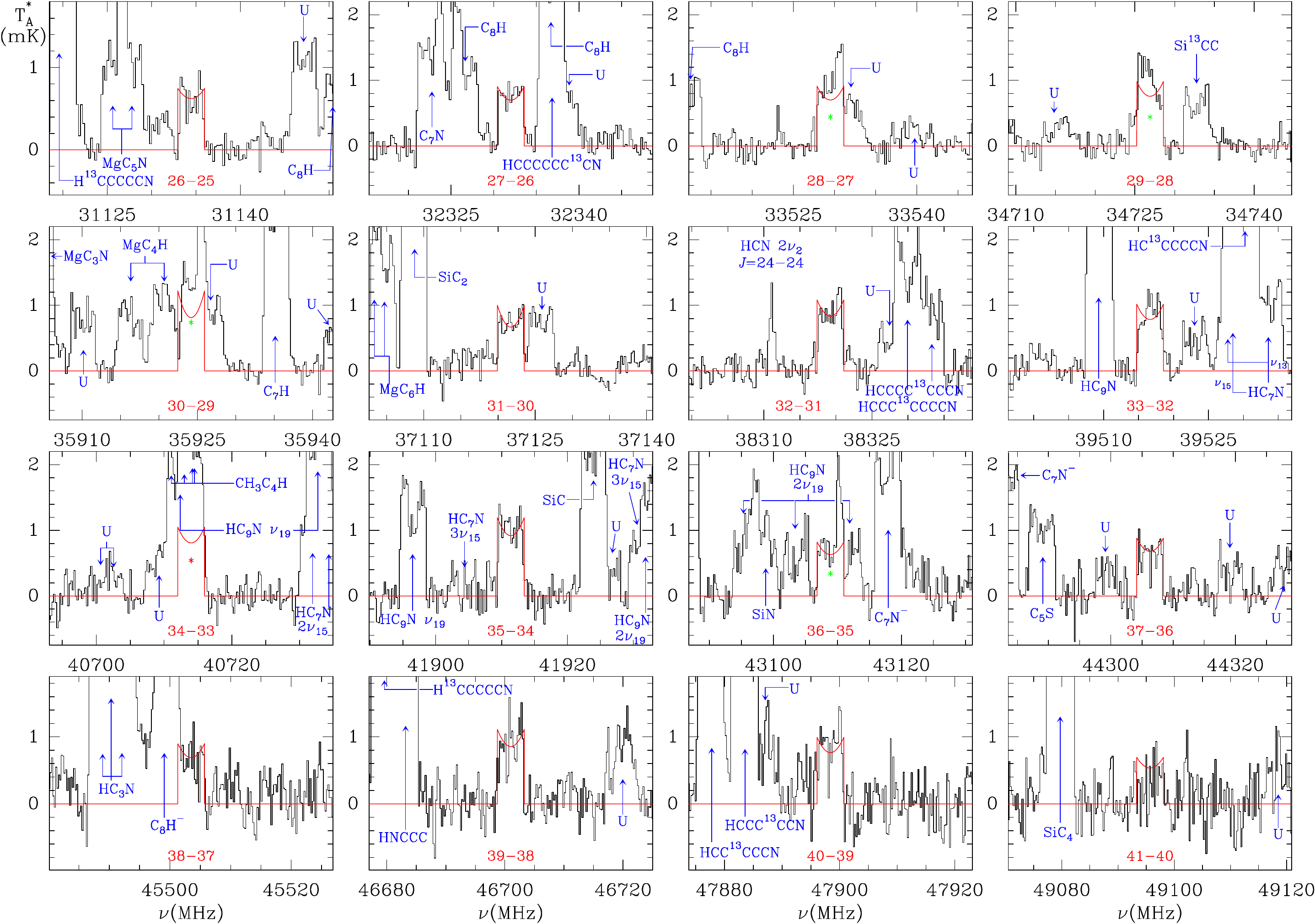}
\caption{Same as Fig. \ref{fig_mgc4h+}, but for the lines of B599 (MgC$_6$H$^+$). 
The line parameters are given in Table \ref{line_parameters_mgc6h+}.
Green stars under the fitted line profile indicate lines marginally blended
with other features but for which the frequency can still be estimated. Red stars indicate lines
heavily blended with other features and for which the frequency cannot be estimated.}
\label{fig_mgc6h+}
\end{figure*}

Ab initio calculations (see Appendix \ref{abinitio} and Table \ref{rota}) indicate
that HCCCNCC and HCCNCCC, both isomers of HC$_5$N, could fit  B1449 and B1447 rather well. 
However, HC$_4$NC is detected
in TMC-1, but not in IRC\,+10216 \citep{Cernicharo2020a}. Hence, it is difficult to 
argue that the carriers of B1449 and B1447, both detected in IRC\,+10216 but
not in TMC-1, 
are HCCCNCC or
HCCNCCC. Moreover, the centrifugal distortion constants for B1449 and B1447 is a factor of  two
larger than those of HC$_5$N, HC$_4$NC, and C$_5$N$^-$, which have a rotational constant 
similar to that of our series (see Table \ref{rota}).
This  points toward two metal-bearing species. MgC$_3$N, AlC$_3$N, NaC$_3$N, and MgC$_4$H have rotational 
and distortion constants very similar to that of B1449 and B1447 (see Appendix \ref{abinitio} and Table \ref{rota}). 
MgC$_3$N and MgC$_4$H, which are $^2\Sigma$ molecules with $\gamma\sim$4.3 MHz, have been identified in 
IRC\,+10216 \citep{Cernicharo2019b}.
Some of the isomers of these two species, MgCCNC, MgNCCC, MgCNCC, and HMgC$_4$, could fit the value of 
the rotational and distortion constants of B1449
and/or B1447 (see Table \ref{rota}), but they   have a value for $\gamma$ similar to that of 
MgC$_3$N, and hence their rotational transitions   appear as well-resolved doublets of constant separation in our observations. The same argument applies to the isotopologs of MgC$_3$N and MgC$_4$H. The vibrationally excited states of these species will exhibit
a $^2\Pi$ vibronic character and cannot be associated with B1449 and B1447.
Some molecules containing sulfur and silicon could also match our series. However, most of them have a $^2\Pi$ ground electronic state and are not detected in IRC\,+10216
(e.g., SiCCCN, SiC$_4$H, and HCCCCS; see Table \ref{rota}).
Some cations of these
species such as SiC$_3$N$^+$ and SiC$_4$H$^+$ have rotational constants close
to that of our series. However, they differ by several MHz (0.5\%; see Table \ref{rota}), and the
calculated distortion constants are too low compared with those of B1449 and B1447. Hence, they
can be also discarded.
C$_4$S has been detected toward TMC-1 \citep{Cernicharo2021b}, but not toward IRC\,+10216. SiC$_4$ 
has $B$=1533.8 MHz (see Table \ref{rota}) and a $^1\Sigma$ ground state. Its lines are prominent in the 
Q-band data of IRC\,+10216, with intensities in the range 20--30 mK. Hence, $^{29}$SiC$_4$
and $^{30}$SiC$_4$ could roughly fit the observed intensities. However, their 
rotational constants have been derived in the
laboratory by \citet{Gordon2000} and do not match those of B1449 and B1447.

\begin{table*}
\small
\caption{New series of lines in IRC\,+10216}
\label{series}
\centering
\begin{tabular}{lcccccccccc}
\hline \hline
 Series   &    $B$                &  $D$           & Carrier      & $B_{calc}$\,$^a$ & $D_{calc}$\,$^a$ & $\mu$& $E_{up}$ range & $T_{rot}$    &   $N$ & $N$(neutral)\,$^b$ \\
          &   (MHz)               & (Hz)           &              &  (MHz)     & (Hz)  &(D) &   (K)        &    (K)       & (10$^{11}$ cm$^{-2}$)&(10$^{11}$ cm$^{-2}$)\\
\hline                                                             
B1449     &  1448.5994$\pm$0.0013 & 63.45$\pm$1.15 & MgC$_4$H$^+$ & 1447.01& 63.6 &13.5 & 9.2-78.0     & 17.4$\pm$0.3 & 4.8$\pm$0.3& 220$\pm$5\,$^c$\\ 
B1447     &  1446.9380$\pm$0.0098 & 91.00$\pm$23.0 & MgC$_3$N$^+$ & 1445.72& 64.2 &18.5 & 9.2-21.2     &  9.1$\pm$0.8 & 1.2$\pm$0.2&  93$\pm$16\,$^c$\\
B599      &   598.7495$\pm$0.0011 &  6.13$\pm$0.43 & MgC$_6$H$^+$ & 598.20 & 6.6  &18.2 & 20.2-49.5    & 21.5$\pm$1.0 & 2.5$\pm$0.3& 200$\pm$90\,$^d$\\
B594      &   594.3176$\pm$0.0026 &  4.92$\pm$1.16 & MgC$_5$N$^+$ & 593.93 & 6.6  &23.7 & 21.6-46.7    & 19.2$\pm$1.6 & 1.1$\pm$0.2&  47$\pm$13\,$^d$\\
\hline
\end{tabular}
\tablefoot{
\tablefoottext{a}{Calculated ab initio rotational and distortion constants (see Appendix~\ref{abinitio} and Table \ref{abini_1}).} 
\tablefoottext{b}{Column density of the corresponding neutral species (MgC$_4$H, MgC$_3$N, MgC$_6$H, or MgC$_5$N).} 
\tablefoottext{c}{\citet{Cernicharo2019b}.} 
\tablefoottext{d}{\citet{Pardo2021}.} 
}
\end{table*}
\normalsize

Among the possible metal-bearing species, AlCCCN and NaCCCN have a $^1\Sigma$ electronic ground state and have been observed in the laboratory
\citep{Cabezas2014,Cabezas2019}. Although their rotational constants do not fit B1449 and B1447, 
some of their isomers, such as 
AlCNCC and AlNCCC or NaNCCC and NaCNCC, could match our series of lines. However, AlCCCN has 
not yet been detected in our data. 
NaCCCN has been recently reported by \citet{Cabezas2023}, but its lines are very weak and its possible isomers 
have rotational constants that do not agree with those of our series (see Table \ref{rota}).
Molecules such as CaC$_4$H, CaC$_3$N, KC$_4$H, and KC$_3$N    produce rotational constants below 1000 MHz \citep{Cabezas2019}.
MgC$_4$ or MgC$_4$$^-$ could fit B1449 or B1447 (see Table \ref{rota}). 
MgC$_4$ is predicted to have a cyclic structure for its ground electronic state
\citep{Redondo2003a}. Nevertheless, it has a linear $^3\Pi$ electronic state that could fit our 
series of lines. We estimated a spin-orbit constant of $-$14.5 cm$^{-1}$ and a 
rotational constant of 1427 MHz (see Appendix \ref{abinitio}), which do not fit
either B1449 or B1447. 
The electronic ground state of MgC$_4$$^-$ is $^2\Sigma$ and its $\gamma$ 
is estimated to be 3-4 MHz producing well-resolved doublets for each rotational transition, which are
not seen in our data.
The anions MgC$_3$N$^-$ and MgC$_4$H$^-$ could be present as the corresponding neutrals have been detected
in IRC\,+10216 \citep{Cernicharo2019b}. However, their rotational constants are 
calculated to be too low to fit B1449 or B1447 (see Table \ref{rota}).

Prior to this work, HCO$^+$ was the only cation known to be present in IRC\,+10216 
\citep{Lucas1990,Agundez2006,
Pulliam2011}. No other cation is predicted to be abundant by chemical models.
However, ab initio calculations by \citet{Baldea2020} and in this work show that MgC$_4$H$^+$ has a $^1\Sigma$ electronic ground state, a dipole 
moment of 13.5\,D, and rotational and distortion constants that fit perfectly those of B1449 (see Tables \ref{series} and \ref{abini_1}). Moreover,
our ab initio calculations for MgC$_3$N$^+$ show that this species also has   a $^1\Sigma$ electronic ground state, a very large dipole
moment of 18.5\,D, and rotational and distortion constants that fit  those of B1447 very well (see Tables \ref{series} and \ref{abini_1}). 
The expected error in the calculated rotational constants is lower than 0.1\% (see Appendix \ref{abinitio}). 
Finally, our ab initio calculations indicate that the rotational
constant of MgC$_4$H$^+$ is always slightly larger than that of MgC$_3$N$^+$ independently of the level
of theory used in the calculations. Hence, the reverse assignment of B1449 to MgC$_3$N$^+$ and
B1447 to MgC$_4$H$^+$ can be ruled out (see Appendix \ref{abinitio}). Consequently, we conclude that the carrier of B1449 is MgC$_4$H$^+$, while that of B1447 is MgC$_3$N$^+$.

\subsection{Detection of B599 (MgC$_6$H$^+$) and B594 (MgC$_5$N$^+$)} \label{det_mgc6h+}

In addition to B1449 and B1447, we found 16 lines in harmonic relation with integer
quantum numbers from $J_u$=26 up to $J_u$=41 in the Q-band data of IRC\,+10216. The lines are shown in Fig. \ref{fig_mgc6h+}. The derived line parameters are given in Table \ref{line_parameters_mgc6h+}. 
The derived rotational constants are $B$=598.7495$\pm$0.0011 MHz and $D$=6.13$\pm$0.43 Hz (B599). The 
lines do not
show any indication of broadening and we checked that dividing the value of $B$ by 2-4 does not match
the observations. Another set of 13 lines were found in harmonic relation with identical quantum numbers
and $B$=594.3176$\pm$0.0026 MHz and $D$=4.92$\pm$1.16 Hz (B594). The lines are shown in Fig. 
\ref{fig_mgc5n+} and
their line parameters are given  in Table \ref{line_parameters_mgc5n+}.
The lines of these two series are not observed in TMC-1.
Hence, similarly to B1449 and B1447, the carriers could have a ground electronic state $^1\Sigma$, or a
$^2\Sigma$ state with $\gamma\le$0.2 MHz. The possible carriers include isomers of HC$_7$N such as 
HC$_5$NC$_2$, HC$_4$NC$_3$, and HC$_3$NC$_4$ (see Table \ref{rota2}).
However, HC$_6$NC
\citep{Botschwina1998} has not yet been  detected in IRC\,+10216, nor in TMC-1. 
Hence, these species can be
discarded as possible carriers of B599 and B594. MgC$_6$ is
also a good potential candidate. However, a cyclic structure is predicted for its ground state
\citep{Redondo2003a}.
Hence, similar arguments to those used
for MgC$_4$ (see Sect. \ref{det_mgc4h+} and Appendix \ref{abinitio})
permit us to discard MgC$_6$ as the carrier of B599 or B594. After considering
all possible carriers, which include isomers of MgC$_5$N and MgC$_6$H among other species, 
we conclude that the best candidate for B599 is MgC$_6$H$^+$, for which
we estimate $B$=598.2 MHz and $D$=6.6 Hz. For B594 we conclude that MgC$_5$N$^+$ is the most 
likely carrier for which our
ab initio calculations provide 
$B$=593.9 MHz and $D$=6.6 Hz (see Table \ref{rota}). 
Similarly to the case of B1449 and B1447, the expected error of the calculated rotational constants of 
MgC$_6$H$^+$ and MgC$_5$N$^+$ is again lower than 0.1\% (see Appendix \ref{abinitio}). 
The calculated dipole moments of MgC$_6$H$^+$ and MgC$_5$N$^+$ are 18.2 and 23.7\,D, 
respectively (see Table \ref{series}).

\section{Discussion} \label{discussion}

The observed spatial distribution of carbon chains observed with ALMA in IRC\,+10216 consists of 
a hollow shell with a radius of 15$''$
\citep{Agundez2017}. Moreover, \cite{Guelin1993} studied the spatial distribution of MgNC and 
concluded 
that this radical also arises from a shell located at 15$''$ from the star. We thus assume here 
that the four new Mg-bearing species 
also arise  from the same region. The cusped shape of the lines argue in favor of this.
From the observed intensities, and through a rotation diagram analysis, we derived the rotational 
temperatures
and column densities for the four species (see Table \ref{series}). 
We note that for MgC$_4$H$^+$, which is the only
species detected in the Q band and in the 3\,mm domain, it is possible to fit the Q-band lines ($J_u\le$17) with a
rotational temperature of 11.9$\pm$0.7\,K, which is  similar to that derived for the 
MgC$_3$N$^+$ lines that cover a similar
energy range. However, including the 3\,mm lines of MgC$_4$H$^+$ results in a rotational temperature of 
17.4$\pm$0.3 K, which is  
similar to the rotational temperatures derived for MgC$_6$H$^+$ and MgC$_5$N$^+$ (see Table \ref{series}).
Taking into account the high dipole moment of the observed species, it seems very unlikely that 
collisional excitation alone could reproduce the high rotational temperatures derived from the data. Infrared
pumping \citep{Cernicharo2014,Pardo2018} may play a role in the population of 
the high-$J$ lines of these molecules.
Infrared pumping is  responsible for the detection of HC$_5$N, HC$_7$N, and HC$_9$N in highly vibrationally excited states with
energies above 100\,K \citep{Pardo2020,Pardo2023}. 

The MgC$_4$H$^+$/MgC$_6$H$^+$ and MgC$_3$N$^+$/MgC$_5$N$^+$ abundance ratios  are, within a factor 
of $\sim$2, similar to those of the corresponding neutral species, MgC$_4$H/MgC$_6$H and 
MgC$_3$N/MgC$_5$N (see Table \ref{series}). However,
the column densities of the cations are 50-80 times lower than those of the corresponding
neutral species (see Table \ref{series}). This difference is likely related to the formation 
mechanism of these Mg-bearing species and the higher reactivity of cations compared to neutrals.

To shed light on the formation of these Mg-bearing cations we modeled the 
chemistry adopting the chemical model presented in \cite{Pardo2021}. We updated the initial abundance of Mg to 3\,$\times$\,10$^{-6}$ relative to H to better reproduce the column densities of Mg-bearing molecules observed in IRC\,+10216. We included the Mg-bearing cations detected here assuming that they are formed by radiative association between Mg$^+$ and the corresponding 
radicals C$_4$H, C$_3$N, C$_6$H, and C$_5$N, while they are mostly destroyed through 
dissociative recombination with electrons and reaction with atomic H. It is unlikely that 
these cations react with H$_2$ because these processes are calculated to be endothermic 
(see Appendix~\ref{energetics}). The most critical parameters are the rate coefficients of 
radiative association between Mg$^+$ and radicals. We therefore performed calculations to 
compute these rate coefficients for Mg$^+$ reacting with C$_4$H, C$_6$H, C$_3$N, and C$_5$N 
(see Appendix~\ref{radiative_association}).

\begin{figure}
\centering
\includegraphics[width=0.48\textwidth]{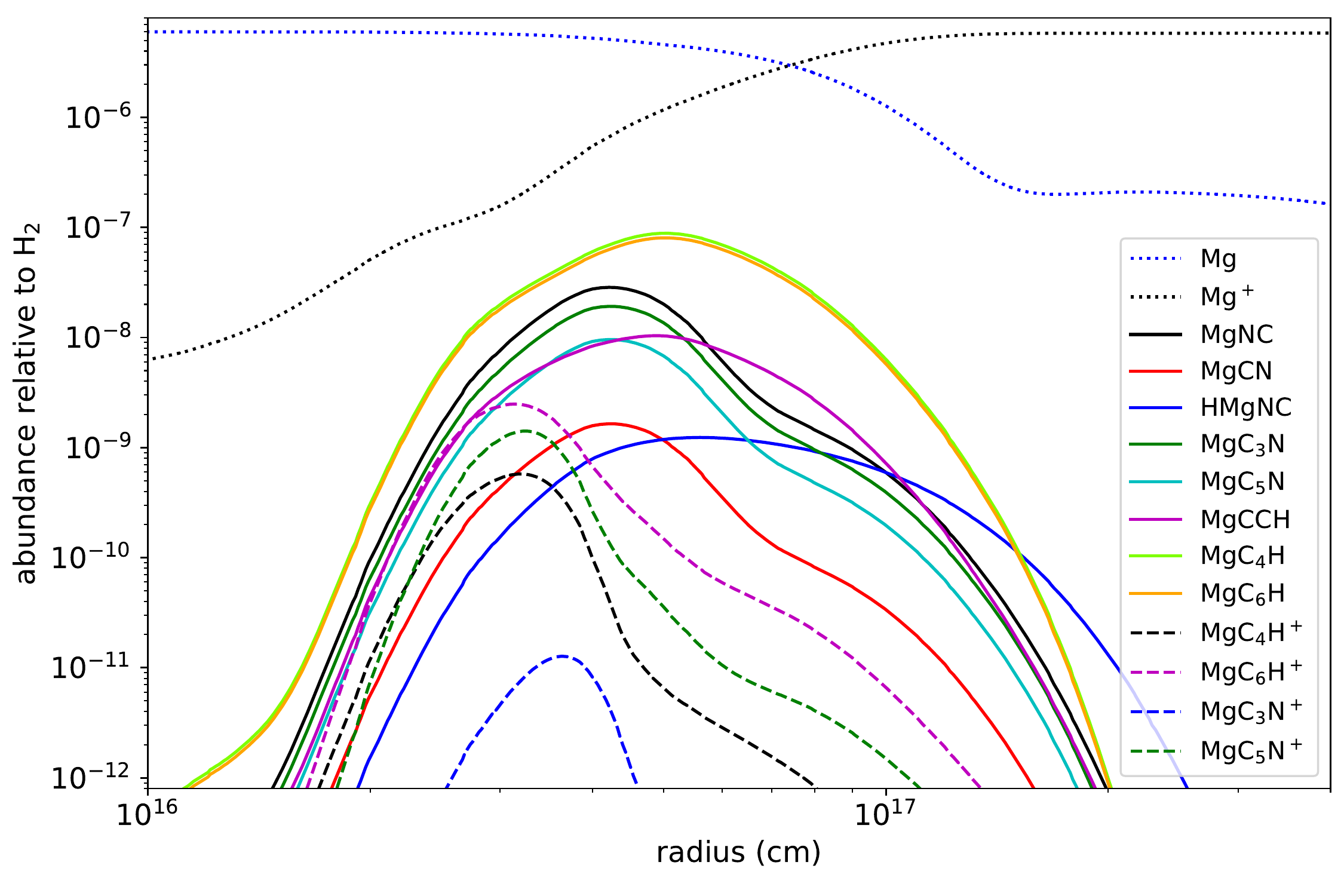}
\caption{Calculated abundances for Mg-bearing species, neutral and
cationic, resulting from the chemical model described in Sect.~\ref{discussion}.}
\label{fig_abun}
\end{figure}

The abundances calculated with our model are shown in Fig. \ref{fig_abun}, where it is seen that Mg-bearing cations are indeed expected to be abundant in the outer layers of IRC\,+10216. 
The model reflects our finding that
the neutral Mg-bearing species are more abundant than their corresponding cations. The column densities predicted for MgC$_4$H$^+$, MgC$_3$N$^+$, MgC$_6$H$^+$, and MgC$_5$N$^+$ are 2.1\,$\times$\,10$^{11}$ cm$^{-2}$, 3.5\,$\times$\,10$^{9}$ cm$^{-2}$, 1.0\,$\times$\,10$^{12}$ cm$^{-2}$, and 4.6\,$\times$\,10$^{11}$ cm$^{-2}$, respectively. The chemical model reproduces the observed column densities within a factor of 4 for the Mg-bearing cations, except for MgC$_3$N$^+$, in which case the calculated column density is about 30 times lower than the observed value (compare with values given in Table~\ref{series}). This is probably due to uncertainties in the rate coefficients of radiative association calculated for the reactions of Mg$^+$ with small molecules such as C$_3$N which has a low-lying $^2\Pi$ electronic excited state. It is also interesting to note that the chemical model predicts that cations reach a peak abundance at smaller distances from the central star compared to the corresponding neutrals. High angular resolution observations of these neutral and positively charged Mg--carbon chains will certainly help to characterize the chemistry of magnesium in IRC\,+10216. Longer Mg-chain cations such as MgC$_8$H$^+$ and MgC$_7$N$^+$ are predicted to have very large dipole moments and reasonable abundances. They could be detected in future improved Yebes\,40m data of IRC\,+10216.
Smaller cationic species such as MgCCH$^+$ and MgNC$^+$ will have very low abundances
because the radiative association rates of Mg$^+$ with CCH and CN are expected to be very low.

\section{Conclusions}

We report in this work the detection of four new series of millimeter-wave lines harmonically 
related in the carbon-rich circumstellar envelope of IRC\,+10216.
From the shape and widths of the lines, which are characteristic of optically thin lines arising in
the outer envelope, and from the absence of nearby spectral features that could reveal associated $K$-components, we conclude that they must arise from four linear molecular species with a $^1\Sigma$ 
electronic ground state.

The sharpness of the U-shaped line edges allows us to derive very accurate values for the spectroscopic
constants of these species. The rotation constants of two species, $B$=1447 and 
$B$=1449 MHz, are similar to that
of cyanodiacetylene, HC$_5$N, a linear C-chain molecule fairly abundant in IRC\,+10216. This implies C-chains 
of similar weight. The distortion constants, $D$, are
twice as large as that of HC$_5$N, which suggests 
metal-bearing species. Similarly, the $B$ and $D$ constants of the two remaining species
($B$=594 and $B$=598 MHz) imply metal-bearing C-chains with a  weight similar to
cyanotriacetylene, HC$_7$N.

From a compilation of the spectroscopic constants of two scores of sulfur, silicon, or metal-bearing C-chains,
their isomers and their ionic forms, derived from laboratory measurements or from our new or previously
published ab initio quantum mechanical calculations, we find a remarkable agreement between $B$ and $D$ of
the four new species and those of the cations MgC$_4$H$^+$, MgC$_3$N$^+$, MgC$_6$H$^+$, and
MgC$_5$N$^+$.

Eight linear Mg-bearing molecules have previously been identified in IRC\,+10216. They include the neutral
species HMgC$_3$N, MgC$_4$H, MgC$_3$N, MgC$_6$H, and MgC$_5$N. Chemical modeling of the outer envelope predict that  
these neutral species
form from the dissociative recombination of MgC$_n$H$_2$$^+$ and MgC$_n$NH$^+$ cations, those cations
being themselves produced by the radiative association of polyynes with Mg$^+$. 
The cations MgC$_n$H$^+$ and MgC$_n$N$^+$ are produced by the radiative association of Mg$^+$ with
the radicals C$_n$H and C$_n$N.
The abundances derived
from the modeling for the neutrals and their cationic species are in 
good agreement with the observed values, giving
confidence that our identifications are correct. This is the first time that cationic metal-bearing
molecular species have been detected in space.

\begin{acknowledgements}
We thank ERC for funding
through grant ERC-2013-Syg-610256-NANOCOSMOS. M.A. thanks MICIU for grant 
RyC-2014-16277. We also thank Ministerio de Ciencia e Innovaci\'on of Spain (MICIU) for funding support through projects
PID2019-106110GB-I00, PID2019-107115GB-C21 / AEI / 10.13039/501100011033, PID2019-106235GB-I00, and PID2021-122549NB-C21.

\end{acknowledgements}

\begin{appendix}
\section{Observations and derived line parameters}
\label{line_parameters}

The observations in the Q band were carried out with the Yebes\,40m radiotelescope.
A detailed description of the telescope, receivers, and backends is 
given by \citet{Tercero2021}. Briefly, the receiver consists of two cold high electron 
mobility transistor amplifiers covering the
31.0-50.3 GHz range (Q band) with horizontal and vertical             
polarizations. Receiver temperatures are 16\,K at 31 GHz and rise to 25\,K at 50 GHz.
The backends are $2\times8\times2.5$ GHz fast Fourier transform spectrometers
with a spectral resolution of 38.15 kHz
providing the whole coverage of the Q band in both polarizations. For the observations of TMC-1
the original spectral resolution was used while for IRC\,+10216 the data were smoothed
to 229 kHz, which corresponds to a velocity resolution 
of 1.7\,km\,s$^{-1}$ at 40 GHz. This spectral resolution
is high enough to resolve the U-shaped lines of IRC\,+10216, which exhibit a full velocity width of
29 km\,s$^{-1}$ \citep{Cernicharo2000}.
The observations of TMC-1 were performed in
frequency-switching mode, while those of IRC\,+10216 were done   in position-switching mode
at the same elevation.
The main beam efficiency of the telescope varies from 0.6 at
32 GHz to 0.43 at 50 GHz.  
The Q-band data of TMC-1 and IRC\,+10216
presented here correspond to 758 h and 696 h, respectively, of on-source telescope 
time. 
All data were analyzed using the GILDAS package.\footnote{\texttt{http://www.iram.fr/IRAMFR/GILDAS}}

\begin{figure*}
\centering
\includegraphics[width=1.0\textwidth]{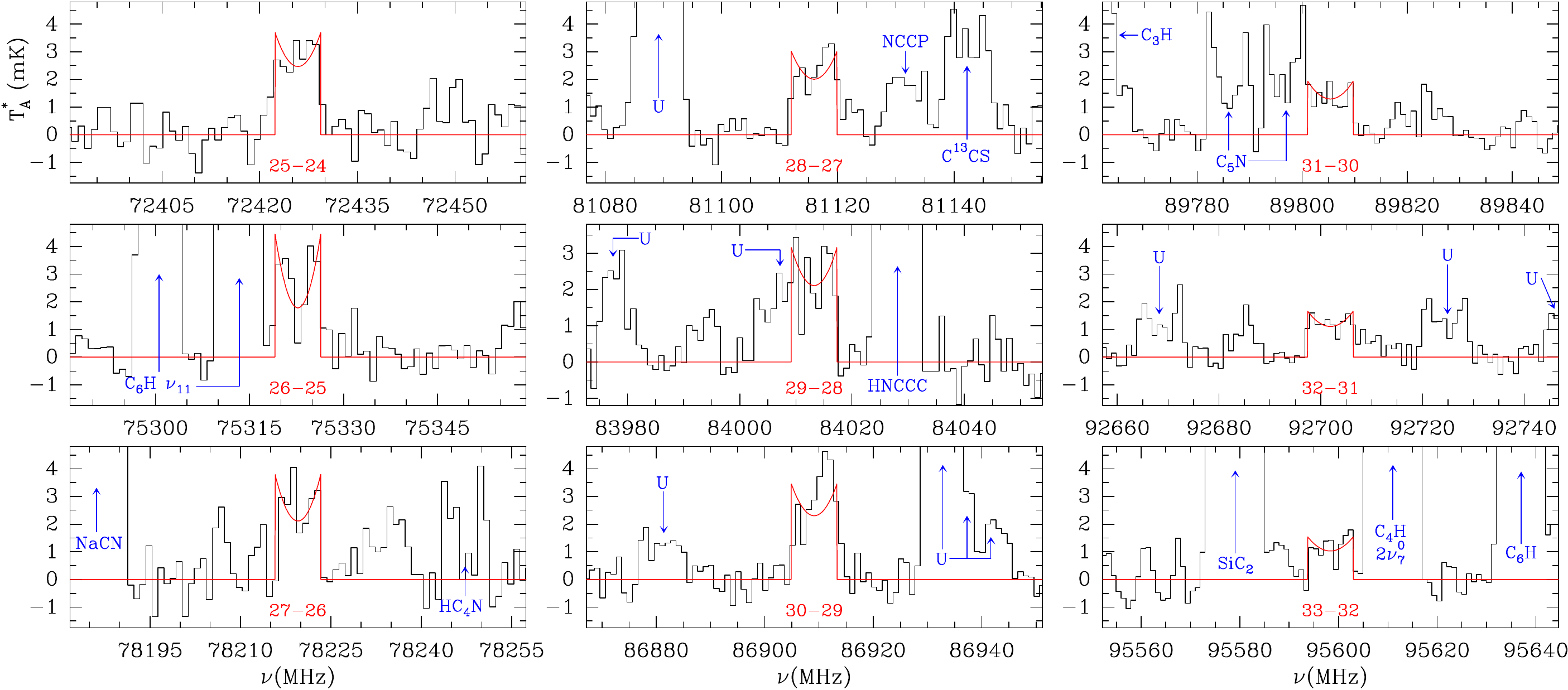}
\caption{Observed lines of B1449 (MgC$_4$H$^+$) with the IRAM\,30m telescope
toward IRC\,+10216. 
Line parameters are given in Table \ref{line_parameters_mgc4h+}.
The abscissa corresponds to the rest frequency assuming a local standard of rest velocity of $-$26.5 \kms.
The ordinate is the antenna temperature corrected for atmospheric and telescope losses in millikelvin.
The red lines show the fitted line profiles.}
\label{fig_mgc4h+_3mm}
\end{figure*}

\begin{figure*}
\centering
\includegraphics[width=1.0\textwidth]{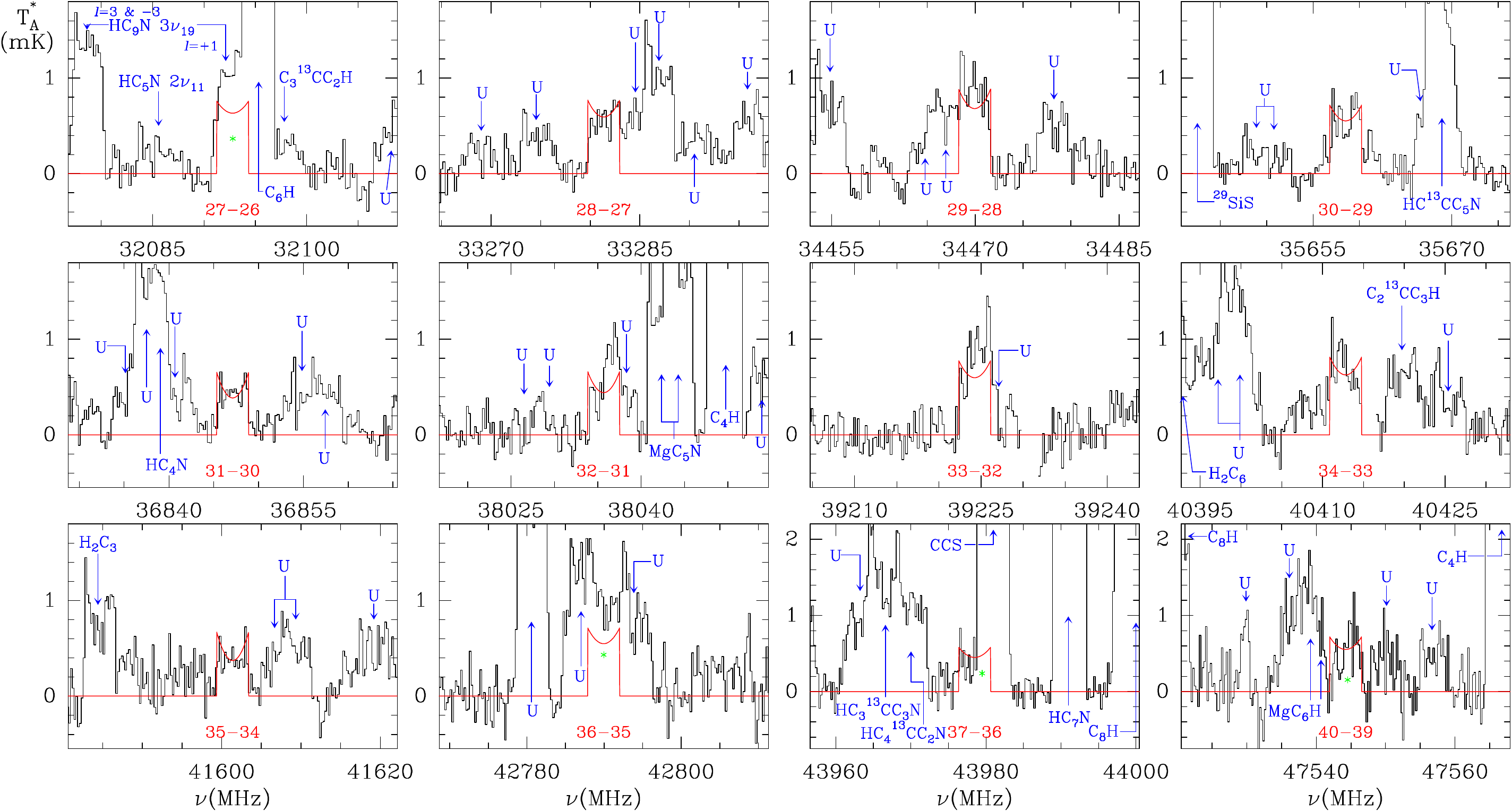}
\caption{Same as Fig. \ref{fig_mgc6h+}, but for the lines of B594 (MgC$_5$N$^+$).
The line parameters are given in Table \ref{line_parameters_mgc5n+}.}
\label{fig_mgc5n+}
\end{figure*}

Further observations at 3\,mm presented in this paper were carried out with the IRAM 30m radiotelescope and 
have been described in detail by \citet{Cernicharo2019a}. 
They correspond to observations acquired during the last 38 years covering the 70--116 GHz 
range and reaching a confidence level (sigma level)
between 0.7 and 3 mK.
Examples of these data can be found in \cite{Cernicharo2019a} 
and \cite{Agundez2014}.
The observing mode, in which we wobbled the secondary mirror by $\pm$90'' at a rate of 0.5 Hz, ensured flat baselines. 

\begin{table}
\tiny
\centering
\caption{Observed line parameters of B1449 (MgC$_4$H$^+$)
toward IRC\,+10216.}
\label{line_parameters_mgc4h+}
\begin{tabular}{cccccccc}
\hline
$J_u$&$\nu_{obs}$$^a$    &$\Delta\nu$$^b$&$\int$ $T_A^*$ dv $^c$ & T$_A^*$\,$^d$ & $\sigma$$^e$&N\\
     & (MHz)             & (MHz)                   &    (mK\,km\,s$^{-1}$)   & (mK)      & (mK)    &\\
\hline
\hline
    11&  31868.88$\pm$0.10 &   0.03& 52.71& 2.32/1.56&0.18&\\
    12&  34765.98$\pm$0.10 &   0.03& 50.90& 2.25/1.50&0.15&\\
    13&  37663.02$\pm$0.10 &  -0.01& 70.08& 3.10/2.07&0.16&\\
    14&  40560.08$\pm$0.10 &  -0.01& 68.03& 3.00/2.50&0.20&\\
    15&  43457.17$\pm$0.10 &   0.04& 69.02& 3.05/2.03&0.22&\\
    16&  46354.02$\pm$0.10 &  -0.12& 73.11& 3.23/2.16&0.36&\\
    17&  49251.16$\pm$0.10 &   0.03& 61.50& 2.72/1.82&0.32&\\
    25&  72425.87$\pm$0.20 &  -0.14& 83.70& 3.69/2.46&0.68&\\
    26&  75322.72$\pm$0.20 &   0.01& 77.81& 4.44/1.78&0.65&\\
    27&  78219.53$\pm$0.20 &   0.16& 77.85& 3.80/2.11&0.67&\\
    28&  81115.95$\pm$0.20 &  -0.05& 68.22& 3.01/2.00&0.67&\\
    29&  84013.25$\pm$0.40 &   0.67& 64.43& 2.84/1.90&0.72&A\\
    30&  86909.10$\pm$0.40 &  -0.01& 62.55& 2.54/1.96&0.72&\\
    31&  89805.40$\pm$0.40 &  -0.20& 43.78& 1.94/1.30&0.72&A\\
    32&  92701.90$\pm$0.40 &  -0.15& 37.52& 1.66/1.11&0.71&\\
    33&  95598.35$\pm$0.40 &  -0.09& 34.80& 1.54/1.03&0.75&\\
\hline  
\hline                                                                                             
\end{tabular}
\tablefoot{\\
\tablefoottext{a}{Observed frequency assuming a v$_{LSR}$ of $-$26.5 \kms.}
\tablefoottext{b}{Observed minus calculated frequencies in MHz.}
\tablefoottext{c}{Integrated line intensity in mK\,km\,s$^{-1}$. The total uncertainty is assumed to be
dominated by the calibration uncertainties of 10\%.}
\tablefoottext{d}{Antenna temperature at the terminal velocity (horn) and at line 
center in millikelvin.}
\tablefoottext{e}{Root mean square noise of the data.}
\tablefoottext{A}{Line partially blended with another feature. Frequency 
and line parameters can still be estimated.}
}
\end{table}
\normalsize

\begin{table}
\tiny
\centering
\caption{Observed line parameters of B1447 (MgC$_3$N$^+$)
toward IRC\,+10216.}
\label{line_parameters_mgc3n+}
\begin{tabular}{cccccccc}
\hline
$J_u$&$\nu_{obs}$$^a$    &$\Delta\nu$$^b$&$\int$ $T_A^*$ dv $^c$ & T$_A^*$\,$^d$ & $\sigma$$^e$&N\\
     & (MHz)             & (MHz)                   &    (mK\,km\,s$^{-1}$)   & (mK)      & (mK)    &\\
\hline
\hline
    11&  31832.13$\pm$0.20 &  -0.02& 25.48& 1.12/0.75&0.17&A\\
    12&  34725.98$\pm$0.20 &   0.10& 17.30& 0.76/0.51&0.16&A\\
    13&  37619.57$\pm$0.10 &  -0.02& 19.96& 0.88/0.69&0.16&\\
    14&  40513.38$\pm$0.10 &   0.11& 24.47& 1.08/0.73&0.21&\\
    15&  43406.73$\pm$0.10 &  -0.18& 23.53& 1.04/0.69&0.23&A\\
    16&  46300.70$\pm$0.20 &   0.17& 22.06& 0.97/0.65&0.36&A\\
    17&  49194.21$\pm$0.20 &   0.11& 18.32& 0.81/0.54&0.33&A\\
\hline  
\hline                                                                                             
\end{tabular}
\tablefoot{\\
\tablefoottext{a}{Observed frequency assuming a v$_{LSR}$ of $-$26.5 \kms.}
\tablefoottext{b}{Observed minus calculated frequencies in MHz.}
\tablefoottext{c}{Integrated line intensity in mK\,km\,s$^{-1}$. The total uncertainty is assumed to be
dominated by the calibration uncertainties of 10\%.}
\tablefoottext{d}{Antenna temperature at the terminal velocity (horn) and at 
line center in millikelvin.}
\tablefoottext{e}{Root mean square noise of the data.}
\tablefoottext{A}{Line partially blended with another feature. Frequency 
and line parameters can still be estimated.}

}
\end{table}
\normalsize

\begin{table}
\tiny
\centering
\caption{Observed line parameters of B599 (MgC$_6$H$^+$)
toward IRC\,+10216.}
\label{line_parameters_mgc6h+}
\begin{tabular}{cccccccc}
\hline
$J_u$&$\nu_{obs}$$^a$    &$\Delta\nu$$^b$&$\int$ $T_A^*$ dv $^c$ & T$_A^*$\,$^d$ & $\sigma$$^e$&N\\
     & (MHz)             & (MHz)                   &    (mK\,km\,s$^{-1}$)   & (mK)      & (mK)    &\\
\hline
\hline
    26&  31134.45$\pm$0.10&-0.09& 19.31 &0.74/0.62&0.14&\\
    27&  32332.00$\pm$0.10& 0.01& 22.31 &0.90/0.70&0.16& \\
    28&  33529.50$\pm$0.15& 0.07& 22.31 &0.91/0.70&0.13&A\\
    29&  34726.90$\pm$0.15& 0.03& 24.14 &0.98/0.76&0.15&A\\
    30&  35924.19$\pm$0.15&-0.11& 27.62 &1.22/0.81&0.17&A\\
    31&  37121.71$\pm$0.10&-0.03& 22.68 &1.00/0.67&0.17& \\
    32&  38319.23$\pm$0.10& 0.07& 26.67 &1.08/0.84&0.16& \\
    33&  39516.64$\pm$0.10& 0.06& 25.10 &1.01/0.78&0.22& \\
    34&  40714.00$\pm$0.02& 0.00&       &         &0.18&B\\
    35&  41911.40$\pm$0.10&-0.01& 29.40 &1.19/0.92&0.19& \\
    36&  43108.90$\pm$0.15& 0.08& 20.22 &0.82/0.63&0.22&A\\
    37&  44306.30$\pm$0.10& 0.08& 21.71 &0.88/0.68&0.27& \\
    38&  45503.50$\pm$0.10&-0.11& 22.13 &0.90/0.69&0.35& \\
    39&  46701.00$\pm$0.10&-0.01& 27.36 &1.11/0.86&0.32& \\
    40&  47898.40$\pm$0.10& 0.01& 24.48 &0.99/0.77&0.32& \\
    41&  49095.70$\pm$0.20&-0.07& 17.14 &0.70/0.54&0.35& \\
\hline  
\hline                                                                                             
\end{tabular}
\tablefoot{\\
\tablefoottext{a}{Observed frequency assuming a v$_{LSR}$ of $-$26.5 \kms.}
\tablefoottext{b}{Observed minus calculated frequencies in MHz.}
\tablefoottext{c}{Integrated line intensity in mK\,km\,s$^{-1}$. The total uncertainty is assumed to be
dominated by the calibration uncertainties of 10\%.}
\tablefoottext{d}{Antenna temperature at the terminal 
velocity (horn) and at line center in millikelvin.}
\tablefoottext{e}{Root mean square noise of the data.}
\tablefoottext{A}{Line partially blended with another feature. Frequency 
and line parameters can still be estimated.}
\tablefoottext{B}{Line heavily blended with another feature. Frequency 
and line parameters cannot be estimated. The quoted frequency corresponds to the
predicted value.}
}
\end{table}
\normalsize

\begin{table}
\tiny
\centering
\caption{Observed line parameters of B594 (MgC$_5$N$^+$)
toward IRC\,+10216.}
\label{line_parameters_mgc5n+}
\begin{tabular}{cccccccc}
\hline
$J_u$&$\nu_{obs}$$^a$    &$\Delta\nu$$^b$&$\int$ $T_A^*$ dv $^c$ & T$_A^*$\,$^d$ & $\sigma$$^e$&N\\
     & (MHz)             & (MHz)                   &    (mK\,km\,s$^{-1}$)   & (mK)      & (mK)    &\\
\hline
\hline
    27&  32092.79$\pm$0.20& 0.03&19.61 &0.76/0.63 &0.16&A\\
    28&  33281.34$\pm$0.15&-0.01&18.91 &0.77/0.59 &0.17&A\\
    29&  34469.89$\pm$0.15&-0.05&21.74 &0.89/0.68 &0.14&A\\
    30&  35658.56$\pm$0.10& 0.04&17.65 &0.72/0.55 &0.14& \\
    31&  36847.17$\pm$0.10& 0.06&13.83 &0.66/0.39 &0.14& \\
    32&  38035.77$\pm$0.15& 0.09&14.97 &0.66/0.44 &0.18&A\\
    33&  39224.30$\pm$0.15& 0.05&19.10 &0.78/0.60 &0.15&A\\
    34&  40412.70$\pm$0.10&-0.12&20.02 &0.82/0.63 &0.19& \\
    35&  41601.31$\pm$0.10&-0.07&13.69 &0.67/0.37 &0.20& \\
    36&  42789.84$\pm$0.15&-0.11&17.51 &0.71/0.55 &0.19&A\\
    37&  43978.60$\pm$0.30& 0.10&14.31 &0.58/0.45 &0.20&A\\
    38&  45167.40$\pm$0.20& 0.35&17.92 &0.73/0.56 &0.33&A\\
    39&  46355.60$\pm$0.09&            &     &    &0.36&B\\
    40&  47544.20$\pm$0.30& 0.06&17.65 &0.72/0.55 &0.40&A\\
    41&  48732.68$\pm$0.12&            &     &    &0.36&C\\
    42&  49921.22$\pm$0.14&            &     &    &1.66&C\\
\hline  
\hline                                                                                             
\end{tabular}
\tablefoot{\\
\tablefoottext{a}{Observed frequency assuming a v$_{LSR}$ of $-$26.5 \kms.}
\tablefoottext{b}{Observed minus calculated frequencies in MHz.}
\tablefoottext{c}{Integrated line intensity in mK\,km\,s$^{-1}$. The total uncertainty is assumed to be
dominated by the calibration uncertainties of 10\%.}
\tablefoottext{d}{Antenna temperature at the terminal velocity (horn) and at line 
center in millikelvin.}
\tablefoottext{e}{Root mean square noise of the data.}
\tablefoottext{A}{Line partially blended with another feature. Frequency 
and line parameters can still be estimated.}
\tablefoottext{B}{Line heavily blended with another feature. Frequency 
and line parameters cannot be estimated. The quoted frequency corresponds to the
predicted one.}
\tablefoottext{C}{Line not detected. The quoted frequency corresponds to the
predicted value.}
}
\end{table}
\normalsize

The intensity scale used in this work, antenna temperature
($T_A^*$), was calibrated using two absorbers at different temperatures and the
atmospheric transmission model ATM \citep{Cernicharo1985, Pardo2001}.
Calibration uncertainties were adopted to be 10~\% for both the IRAM 30m data and the Yebes 40m data.
Additional uncertainties could arise in the case of IRC\,+10216 from the line intensity 
fluctuation with time induced by the stellar
infrared flux variability \citep{Cernicharo2014,Pardo2018}.
Pointing corrections were obtained by observing strong
nearby quasars and the SiO masers of R\,Leo. Pointing errors
were always within 2-3$''$.

We used the SHELL method of the CLASS program, within the GILDAS package, which is well adapted for the line profiles observed in circumstellar envelopes. A variable window of 50--100 MHz was used to
derive the line parameters in this source, which are given in Tables 
\ref{line_parameters_mgc4h+}, \ref{line_parameters_mgc3n+},
\ref{line_parameters_mgc6h+}, and \ref{line_parameters_mgc5n+}
for MgC$_4$H$^+$, MgC$_3$N$^+$, MgC$_6$H$^+$, and MgC$_5$N$^+$, respectively. 
The Q-band lines of MgC$_4$H$^+$ observed with the Yebes\,40m telescope are shown in 
Fig.~\ref{fig_mgc4h+}, while those observed at higher frequencies
with the IRAM\,30m telescope are shown in Fig. \ref{fig_mgc4h+_3mm}. The lines of MgC$_3$N$^+$,
MgC$_6$H$^+$, and MgC$_5$N$^+$ are shown 
in Figures \ref{fig_mgc4h+} (right panels), \ref{fig_mgc6h+}, and \ref{fig_mgc5n+}, respectively. 
In all cases we fixed the terminal expansion velocity
for the circumstellar envelope to 14.5 \kms \citep{Cernicharo2000}. Most lines 
of MgC$_4$H$^+$ are free of blending. Two of
them ($J$=29-28 and $J$=31-30) have a marginal overlap with another feature,
but a fit to the line profile is still possible. For the other species several lines are 
partially blended, but it is still possible to derive their frequencies (lines indicated in Figures
\ref{fig_mgc6h+} and \ref{fig_mgc5n+} with a green star). Only the $J$=34-33 line of MgC$_6$H$^+$ is
heavily blended (red star in Figure \ref{fig_mgc6h+}).

\section{Ab initio calculations}
\label{abinitio}

We performed ab initio calculations to obtain accurate rotational parameters for all the candidates to be carriers of the series of harmonically related lines found. We also carried out calculations for analogous molecular species for which the rotational parameters are experimentally 
known. In this manner we were able to scale the calculated values for the target species using 
experimental-to-theoretical ratios derived from the related species.

\begin{table*}
\caption{Spectroscopic parameters $B$ (MHz) and $D$ (Hz) for MgC$_4$H$^+$, MgC$_3$N$^+$, MgC$_6$H$^+$, and MgC$_5$N$^+$ species.}
\label{abini_1}
\centering
\begin{tabular}{{lcccccc}}
\hline
\hline
&\multicolumn{2}{c}{MgC$_4$H}&\multicolumn{3}{c}{MgC$_4$H$^+$} \\
\cmidrule(lr){2-3} \cmidrule(lr){4-6}
Parameter & Calc.\tablefootmark{a} & Exp.\tablefootmark{a} & Calc.\tablefootmark{b} & Scaled\tablefootmark{b} & Exp.\tablefootmark{b}\\
\hline
$B$            &   1377.38     &     1381.512   &  1442.68   &  1447.01     &  1448.5994(13)     \\
$D$            &  60.1         &   74.0         &  51.6      &  63.6        &     63.45(115)      \\
\hline
\hline
&\multicolumn{2}{c}{MgC$_3$N}&\multicolumn{3}{c}{MgC$_3$N$^+$} \\
\cmidrule(lr){2-3} \cmidrule(lr){4-6}
Parameter & Calc.\tablefootmark{a} & Exp.\tablefootmark{a} & Calc.\tablefootmark{b} & Scaled\tablefootmark{b} & Exp.\tablefootmark{b}\\
\hline
$B$            &    1376.50   &     1380.888    &  1441.12   & 1445.72     &  1446.9380(98)     \\
$D$            &  64.0        &   76.0          &  54.4      &  64.2       &    91(23)   \\
\hline
\hline
&\multicolumn{2}{c}{MgC$_6$H}&\multicolumn{3}{c}{MgC$_6$H$^+$} \\
\cmidrule(lr){2-3} \cmidrule(lr){4-6}
Parameter & Calc.\tablefootmark{c} & Exp.\tablefootmark{c} & Calc.\tablefootmark{b} & Scaled\tablefootmark{b} & Exp.\tablefootmark{b}\\
\hline
$B$            &    578.11   &     579.75150    &  596.51    & 598.20     &  598.7495(11)     \\
$D$            &  5.83        &   7.325\tablefootmark{d}   &  5.24      &  6.58       &    6.13(43)   \\
\hline
\hline
&\multicolumn{2}{c}{MgC$_5$N}&\multicolumn{3}{c}{MgC$_5$N$^+$} \\
\cmidrule(lr){2-3} \cmidrule(lr){4-6}
Parameter & Calc.\tablefootmark{c} & Exp.\tablefootmark{c} & Calc.\tablefootmark{b} & Scaled\tablefootmark{b} & Exp.\tablefootmark{b}\\
\hline
$B$            &   574.74      &     576.43     &    592.19   &  593.93    &  594.3176(26)     \\
$D$            &  5.87         &   7.325        &    5.31      &  6.62     &     4.92(116)      \\
\hline
\hline
\end{tabular}
\tablefoot{\\
  \tablefoottext{a}{\citet{Cernicharo2019b}.} 
\tablefoottext{b}{This work.} 
\tablefoottext{c}{\citet{Pardo2021}.}
\tablefoottext{d}{Not well determined in \citet{Pardo2021} and assumed to be the same than that for MgC$_5$N.}
}
\end{table*}

The geometry optimization calculations for all the species were done using the coupled cluster 
method with single, double, and perturbative triple excitations with an explicitly correlated 
approximation (CCSD(T)-F12; \citealt{Knizia2009}) and all electrons (valence and core) correlated 
together with the Dunning's correlation consistent basis sets with polarized core-valence 
correlation triple-$\zeta$ for explicitly correlated calculations (cc-pCVTZ; \citealt{Hill2010}). 
At the optimized geometry, the electric dipole moment was calculated at the same level of theory as that 
for the geometrical optimization. These calculations were carried out using the Molpro 2020.2 program 
\citep{Werner2020}. The values for the centrifugal distortion constants were obtained using harmonic 
vibrational frequency calculations, with the second-order M{\o}ller-Plesset perturbation 
theory method (MP2; \citealt{Moller1934}) and the correlation consistent with polarized valence 
triple-$\zeta$ basis set (cc-pVTZ; \citealt{Woon1993}). These calculations were carried out using 
the Gaussian16 program \citep{Frisch2016}.

Table \ref{abini_1} shows the results obtained for MgC$_4$H$^+$, MgC$_3$N$^+$, MgC$_6$H$^+$, 
and MgC$_5$N$^+$, while the dipole moments are shown in Table \ref{series}. The rotational constant 
$B$ was calculated at CCSD(T)-F12/cc-pCVTZ-F12 level of theory, while the constant $D$ was 
calculated at MP2/cc-pVTZ level of theory. The ab initio values calculated for MgC$_4$H$^+$, MgC$_3$N$^+$, 
MgC$_6$H$^+$, and MgC$_5$N$^+$ are scaled by the experimental-to-calculated ratio of the corresponding parameter 
of MgC$_4$H, MgC$_3$N, MgC$_6$H, and MgC$_5$N, respectively.

We followed the same procedure for the  HCCCNCC and HCCNCCC species, whose scaled rotational parameters, 
using HC$_5$N as reference, are shown in Table \ref{rota}. As for MgC$_4$H$^+$, MgC$_3$N$^+$, MgC$_6$H$^+$, 
and MgC$_5$N$^+$, the rotational constants $B$ were calculated at CCSD(T)-F12/cc-pCVTZ-F12 level of theory, 
while the constants $D$ were calculated at MP2/cc-pVTZ level of theory. 
Table \ref{rota}
also provides the rotational constants of the species discussed in Sect. \ref{det_mgc4h+}.

MgC$_4$ and MgC$_6$ are predicted to have $^1A_1$ cyclic structures as 
electronic ground states \citep{Redondo2003a}. 
Nevertheless, these species
also have high energy linear isomers with $^3\Pi$ electronic ground states.
For the $^3\Pi$ linear isomers of MgC$_4$ and MgC$_6$ 
we carried out the same type of calculations, and  chose MgC$_4$H and MgC$_6$H as 
reference species to scale the ab initio parameters (see Tables \ref{rota} \& \ref{rota2}). 
We estimate a spin-orbit constant $A_{SO}$ of $-$14.5 cm$^{-1}$ for both species. 
Three series of lines corresponding to the $^3\Pi_2$, $^3\Pi_1$, and $^3\Pi_0$ ladders could
be expected, with $^3\Pi_2$ having the lowest rotation constant and the lowest energy levels.
The
rotational constant of linear MgC$_4$ is $\sim$22 MHz below those of B1449 and B1447 so that it can be
ruled out as a possible carrier of these species. The rotational constant of MgC$_6$ is predicted
to be $\sim$597 MHz (see Table \ref{rota2}) and could fit B598 and/or B594. However, we could expect to have other series
of lines corresponding to the $^3\Pi_1$ and $^3\Pi_0$ ladders  with
decreasing intensity that are not found in our data. The possibility that B594 is the $^3\Pi_2$ ladder and
the B598 the $^3\Pi_1$ ladder can be discarded as the lines of B594 have less intensity than those of B598.
Moreover, no other series is found around $B$=602 MHz that could be assigned to the $^3\Pi_0$ ladder.
These arguments, together with the fact that the lowest energy isomers are expected to be cyclic,
permit us to discard MgC$_4$ and MgC$_6$ as carriers of some of our series of lines.

From a comparison between our calculated and experimental values for
species observed in the laboratory, 
we estimate that the predicted rotational constants for all considered molecules have uncertainties 
better than 0.1\%.

\begin{table}
\caption{Rotational constants for potential candidates of B1449 and B1447.}
\label{rota}
\centering
\begin{tabular}{lcccc}
\hline
\hline
 Molecule    &    $B$      &  $D$    & GS & Notes\\
                         &   (MHz)     & (Hz)    &    &         \\
\hline
HC$_5$N      &  1331.3327           & 30.1      & $^1\Sigma$ &A,B,C,3\\
HC$_4$NC     &  1401.1822           & 32.8      & $^1\Sigma$ &B,C,4\\
HCCCNCC             &  1448.3                        & 33.7           & $^1\Sigma$ & 2\\
HCCNCCC             &  1458.1                        & 33.2           & $^1\Sigma$ & 2\\
HNC$_5$      &  1358.3                    &                & $^1\Sigma$ & 4\\
C$_5$N$^-$   &  1388.8673        & 35.7               & $^1\Sigma$ &5 \\
AlCCCN       &  1340.7575        & 67.3               & $^1\Sigma$ &C,6 \\
AlCCNC       &  1395.8           &                    & $^1\Sigma$ &6 \\
AlCNCC       &  1419.7           &                    & $^1\Sigma$ &6 \\
AlNCCC       &  1480.9           &                    & $^1\Sigma$ &6 \\
NaCCCN       &  1321.0943        & 101.1              & $^1\Sigma$ &C,7\\
NaCCNC       &  1399.8           &                    & $^1\Sigma$ &7 \\
NaCNCC       &  1444.2           &                    & $^1\Sigma$ &7 \\
NaNCCC       &  1489.1           &                    & $^1\Sigma$ &7 \\
MgCCCN       &  1380.9975        & 76.70         & $^2\Sigma$ &A,C,8 \\
MgCCNC             &  1469.1/1475      &               & $^2\Sigma$ & 7,9\\
MgNCCC             &  1536.1/1561      &               & $^2\Sigma$ & 7,9\\
MgCNCC       &  1479.2           &               & $^2\Sigma$ & 7\\
CCMgCN       &  1391.7           &               & $^2\Sigma$ & 7\\
CCMgNC       &  1505.4           &               & $^2\Sigma$ & 7\\
MgCCCN$^-$   &  1275.0           &                &$^1\Sigma$ & 9\\
MgNCCC$^-$   &  1414.0           &                &$^1\Sigma$ & 9\\
MgCCNC$^-$   &  1351.0           &                &$^1\Sigma$ & 9\\
MgC$_4$H     &  1381.512         & 73.70          & $^2\Sigma$ &A,C,8 \\
HMgC$_4$     &  1406                       &                & $^2\Sigma$ & 9\\
MgC$_4$H$^-$ &  1281/1288        & 144.5           & $^1\Sigma$ & 1,9\\ 
HMgC$_4$$^-$ &  1420             &                & $^1\Sigma$ & 9\\ 
MgC$_4$      &  1427            &                & $^3\Pi$    &2\\ 
MgC$_4$$^-$  &  1463/1487            &                & $^2\Sigma$ &10\\ 
SiCCCN       & 1414.740         & 58.31          &  $^2\Pi$ & C,11\\
SiC$_3$N$^+$ & 1455.3           & 45.0           &  $^1\Sigma$ & 2\\
SiCCCC       & 1533.772         & 58.36          & $^1\Sigma$  & C,12\\
SiC$_4$H     & 1415.703         & 56.0           &  $^2\Pi$ & C,13\\
SiC$_4$H$^+$ & 1461.6           & 42.0           &  $^1\Sigma$ & 2\\
C$_4$S       & 1519.206         & 48.05          & $^3\Sigma$ &C,14\\
HC$_4$S      & 1434.3255        & 47.00          &  $^2\Pi$ & C,15\\
\hline
\hline
\end{tabular}
\tablefoot{\\
\tablefoottext{A}{Detected in IRC\,+10216.}
\tablefoottext{B}{Detected in TMC-1.}
\tablefoottext{C}{Rotational constants from laboratory and/or astronomical data.}
\tablefoottext{1}{\citet{Baldea2020}}  
\tablefoottext{2}{Ab initio calculations from this work (see Appendix~\ref{abinitio} for details).} 
\tablefoottext{3}{\citet{Giesen2020} and references therein} 
\tablefoottext{4}{\citet{Cernicharo2020a} and references therein} 
\tablefoottext{5}{\citet{Cernicharo2020b}.} 
\tablefoottext{6}{\citet{Cabezas2014}.} 
\tablefoottext{7}{\citet{Cabezas2019}.} 
\tablefoottext{8}{\citet{Cernicharo2019b}.} 
\tablefoottext{9}{\citet{Aoki2012}.} 
\tablefoottext{10}{\citet{Redondo2003b}.} 
\tablefoottext{11}{\citet{Umeki2014}.} 
\tablefoottext{12}{\citet{Ohishi1989,Gordon2000}.} 
\tablefoottext{13}{\citet{McCarthy2001}.} 
\tablefoottext{14}{\citet{Hirahara1993,Gordon2001}.} 
\tablefoottext{15}{\citet{Hirahara1994}.} 

}
\\
\end{table}

\begin{table}
\caption{Rotational constants for potential candidates of B598 and B594.}
\label{rota2}
\centering
\begin{tabular}{lcccc}
\hline
\hline
 Molecule    &    $B$      &  $D$    & GS & Notes\\
                         &   (MHz)     & (Hz)    &    &         \\
\hline
HC$_6$NC     &   582.5203 &  5.4  & $^1\Sigma$ & A,1  \\
HC$_5$NC$_2$ &   598.8    &       & $^1\Sigma$ &  2   \\
HC$_4$NC$_3$ &   607.5    &       & $^1\Sigma$ &  2   \\
HC$_3$NC$_4$ &   609.4    &       & $^1\Sigma$ &  2   \\
HC$_2$NC$_5$ &   605.4    &       & $^1\Sigma$ &  2   \\
HCNC$_6$     &   592.0    &       & $^1\Sigma$ &  2   \\
HNC$_7$      &   578.8    &       & $^1\Sigma$ &  2   \\
MgC$_6$      &   597.2    &       & $^3\Pi$    &  2   \\
\hline
\hline
\end{tabular}
\tablefoot{\\
\tablefoottext{A}{Rotational constants from laboratory.}
\tablefoottext{1}{\citet{Botschwina1998}.}
\tablefoottext{2}{Ab initio calculations from this work (see text for details).} 
}
\\
\end{table}

\section{Energetics of the reactions of MgC$_n$H$^+$ and MgC$_n$N$^+$ with H$_2$}
\label{energetics}

We explored, using quantum chemical calculations, the energetics of all possible channels in the reactions of the cations MgC$_4$H$^+$, MgC$_3$N$^+$, MgC$_6$H$^+$, and MgC$_5$N$^+$ with H$_2$. The energetics of all the 
reactions were calculated considering the ZPE of the separated reactants 
and products at the CCSD/aug-cc-pVTZ level of theory. The results are shown 
in Table~\ref{energies}. It can be seen that all the reaction channels are clearly endothermic, 
which means that these cationic species do not react with H$_2$ at low temperatures. Moreover, 
those channels calculated to have a lower endothermicity, such as channel 3 in the reactions 
MgC$_3$N$^+$ + H$_2$ and MgC$_5$N$^+$ + H$_2$, 
are expected to have quite high barriers because the concerted reactions involve the break of two 
bonds and the formation of  two others. These barriers preclude the reactions from taking place.

\begin{table}
\caption{Reaction energies for the reactions of MgC$_4$H$^+$, MgC$_3$N$^+$, MgC$_6$H$^+$, and MgC$_5$N$^+$ with H$_2$, in kcal mol$^{-1}$.}
\label{energies}
\centering
\begin{tabular}{{lll}}
\hline
\hline
\multicolumn{3}{c}{MgC$_4$H$^+$ + H$_2$} \\
\cmidrule(lr){1-3}
Channel & Products  & $\Delta$E\tablefootmark{a}\\
\hline
1           &   MgC$_4$H$_2$$^+$ ($^2$$B$$_2$) + H ($^2$$P$)      &  56.3      \\
2           &   HMgC$_4$H$^+$ ($^{2}\Pi$) + H ($^2$$P$)           &  83.2       \\
3           &   MgH$^+$ ($^{1}\Sigma$) + HC$_4$H  ($^{1}\Sigma$)  &  9.1      \\
4           &   MgH ($^{2}\Sigma$) + HC$_4$H$^+$ ($^{2}\Pi$)      &   82.3      \\
5           &   Mg···HC$_4$H$^+$ ($^{2}A'$) + H ($^2$$P$)         &   34.3    \\
& & \\
\hline
\hline
\multicolumn{3}{c}{MgC$_3$N$^+$ + H$_2$} \\
\cmidrule(lr){1-3}
Channel & Products & $\Delta$E\tablefootmark{a}\\
\hline
1           &   MgC$_3$NH$^+$ ($^2$$B$$_2$) + H ($^2$$P$)      &  74.7      \\
2           &   HMgC$_3$N$^+$ ($^{2}\Pi$) + H ($^2$$P$)           &  100.3       \\
3           &   MgH$^+$ ($^{1}\Sigma$) + HC$_3$N  ($^{1}\Sigma$)  &  3.0      \\
4           &   MgH ($^{2}\Sigma$) + HC$_3$N$^+$ ($^{2}\Pi$)      &   108.8      \\
5           &   HC$_3$NMg$^+$ ($^{2}\Sigma$) + H ($^2$$P$)         &   9.1    \\
& & \\
\hline
\hline
\multicolumn{3}{c}{MgC$_6$H$^+$ + H$_2$} \\
\cmidrule(lr){1-3}
Channel & Products & $\Delta$E\tablefootmark{a}\\
\hline
1           &   MgC$_6$H$_2$$^+$ ($^2$$B$$_2$) + H ($^2$$P$)      &  53.1      \\
2           &   HMgC$_6$H$^+$ ($^{2}\Pi$) + H ($^2$$P$)           &  72.8       \\
3           &   MgH$^+$ ($^{1}\Sigma$) + HC$_6$H  ($^{1}\Sigma$)  &  9.5      \\
4           &   MgH ($^{2}\Sigma$) + HC$_6$H$^+$ ($^{2}\Pi$)      &   68.8      \\
5           &   Mg···HC$_4$H$^+$ ($^{2}A'$) + H ($^2$$P$)         &   33.3    \\
& & \\
\hline
\hline
\multicolumn{3}{c}{MgC$_5$N$^+$ + H$_2$} \\
\cmidrule(lr){1-3}
Channel & Products  & $\Delta$E\tablefootmark{a}\\
\hline
1           &   MgC$_5$NH$^+$ ($^2$$B$$_2$) + H ($^2$$P$)      &  74.3      \\
2           &   HMgC$_5$N$^+$ ($^{2}\Pi$) + H ($^2$$P$)           &  103.0       \\
3           &   MgH$^+$ ($^{1}\Sigma$) + HC$_5$N  ($^{1}\Sigma$)  &  4.6      \\
4           &   MgH ($^{2}\Sigma$) + HC$_5$N$^+$ ($^{2}\Pi$)      &   118.3      \\
5           &   HC$_5$NMg$^+$ ($^{2}\Sigma$) + H ($^2$$P$)         &   9.9    \\
\hline
\hline
\end{tabular}
\tablefoot{\\
\tablefoottext{a}{Energy relative to that of separated reactants.} \\
\\
}
\end{table}

\section{Radiative association of Mg$^+$ with the radicals C$_n$H and C$_n$N}
\label{radiative_association}

Radiative association in collisions of a metallic ion, M$^+$ = Mg$^+$, and a polar molecule, A = C$_n$H or C$_n$N,
can be described by the scheme developed by \citet{Herbst1976}, \cite{Bass1981}, \cite{Herbst1987}, and \cite{Klippenstein1996}: 
\begin{eqnarray}
  M^+ + A \stackrel[K_B]{K_F}\rightleftarrows (MA^+)^* \stackrel{K_E}\longrightarrow MA^+  + h\nu.
\end{eqnarray}
The formation of the initial complex, with rate constant $K_F$, is dominated by the strong attractive ion-electric dipole
interactions, which can be written as
\begin{eqnarray}\label{eq:ion-dipole-interaction}
  V= -{q_{ion} {\bf d}\over 4\pi\epsilon_0} {\cos\theta\over R^2},
\end{eqnarray}
where ${\bf d}$ is the electric dipole of the molecule, $q_{ion}$ is
the charge of the ion, ${\bf R}$ is the vector joining the center of mass of A to the ion M$^+$ (of norm $R$),
with $\cos\theta={\bf R}\cdot{\bf d}/ R d$. In the present case the C$_n$H and C$_n$N molecules are linear
with a permanent electric dipole along the molecular axis.

Once the $(MA^+)^*$ complex is formed, it can either dissociate back,
with a rate constant $K_B$, or emit
a photon to estabilize the $MA^+$ ion, with a radiative constant $K_E$. Thus, the total radiative association
rate constant, $K_{RA}$, is obtained as
\begin{eqnarray}
  K_{RA} (T) = K_F(T)  {K_E(T)\over{K_E(T)+K_B(T)}}
.\end{eqnarray}
$K_F(T)$ is rather fast, especially at low temperature, because of the long-range interactions,
and the $K_{RA}$ rate constant depends strongly on the ratio ${K_E(T)\over{K_E(T)+K_B(T)}}$, which is very sensitive
to the molecular properties.

The calculation of the $ K_F(T),K_B(T),$ and $K_E(T)$  rates is achieved here using
statistical methods similar to those previously used by
\citet{Bass1981}, \citet{Klippenstein1996}, and \citet{Dunbar2002}.
First, the microcanonical rate constants for each individual process are obtained as a function of the collisional energy, $E$. Second,
the thermal rate constants are obtained by averaging over $E$ using a Boltzmann distribution. The method
is briefly described here.

\begin{figure}[t]
\begin{center}
  \includegraphics[width=8.8cm]{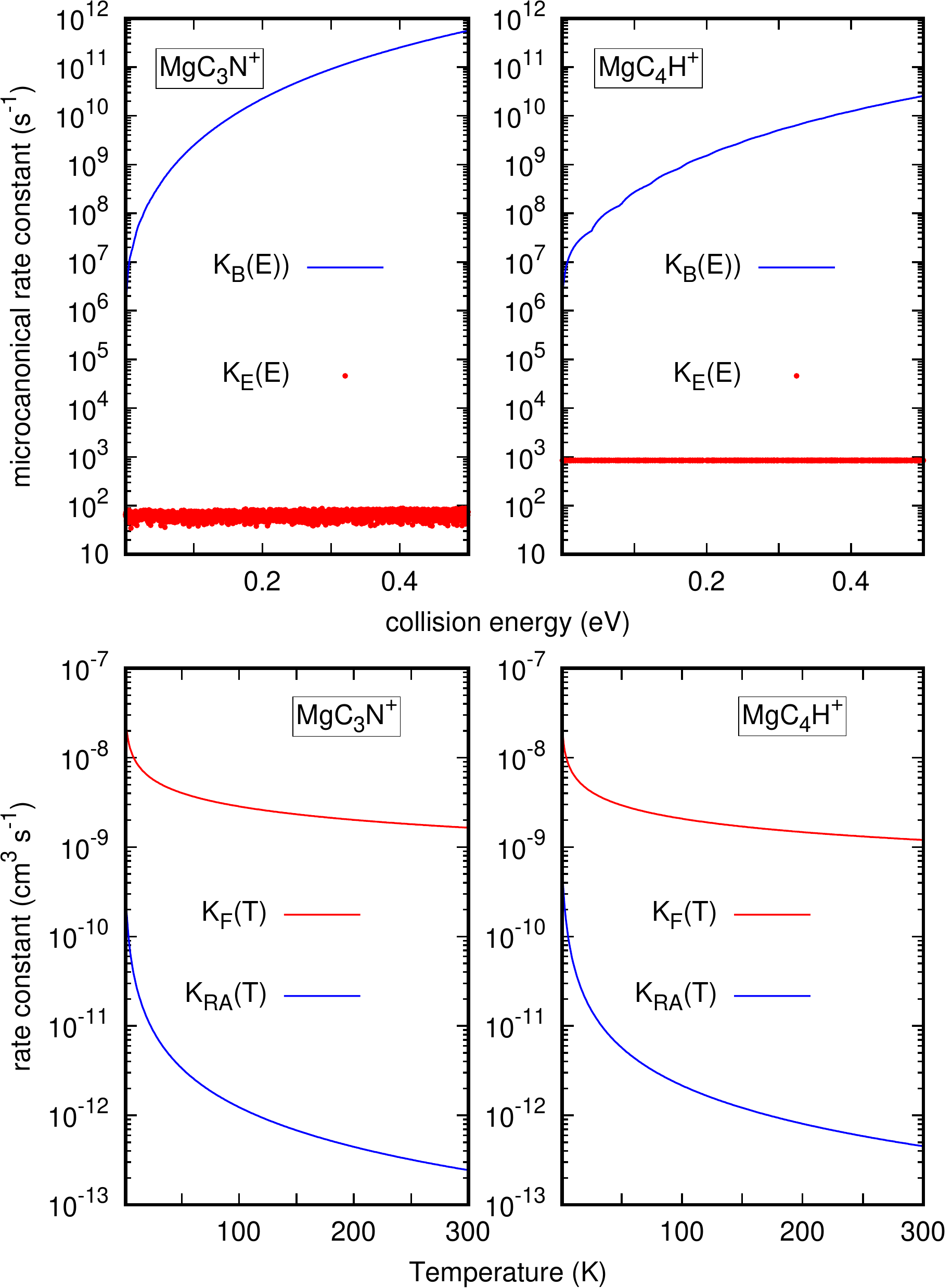}
  
  \caption{\label{fig:rates}{
  Calculated rates for the radiative association of Mg$^+$
  with C$_3$N and C$_4$H.  
  Left panels: Thermal rate constant of formation, $K_F(T)$, and radiative association, $K_{RA}$,
      for MgC$_3$N$^+$ and MgC$_4$H$^+$, as indicated in each panel.
      Right panels: Microcanonical rate constants for the back-dissociation, $K_B(E)=\sum_J(2J+1)*K_B(E,J)$, and the radiative
      emission averaged in each energy interval, $K_E(E)$. 
  }}
\vspace*{0.1cm}
\end{center}
\end{figure}

According to {\citet{Klippenstein1996}}, the microcanonical rate constant
at a given collisional energy, $E$, and total angular momentum, $J$, can be expressed
as
\begin{eqnarray}
  K_X(E,J) ={N(E,J)\over h \rho_X(E,J) }
,\end{eqnarray}
where  $X=F,B$ for formation and back-dissociation, respectively. In this expression,
$N(E,J)$ is the cumulative reaction probability \citep{Klippenstein1996}
and $\rho_F(E,J)$ and $\rho_B(E,J)$ are the density of states of M$^+$ + A  reactants and
of the (MA$^+$)$^*$ complex, respectively. These densities are rather different and play
an important role.

The strong ion-dipole interaction in Eq.~(\ref{eq:ion-dipole-interaction})
rapidly reorients the system to a collinear geometry, and hence a simple
capture model based on the mono-dimensional radial coordinate can be done, similar
to the Langevin model,   obtaining a thermal rate constant as
\begin{eqnarray}
  K_F(T)= q_e \sqrt{ {8 \pi C^2 \over \mu k_B T}}
,\end{eqnarray}
where $q_e=1/4$ is the electronic partition function for C$_3$N, C$_4$H, and C$_5$N (for two doublet reactants
to form a singlet-state complex, all being $\Sigma$ states), and $q_e=1/8$ for C$_6$H. In this expression
$C={q_{ion} {\bf d}/ 4\pi\epsilon_0}$,
which depends on the electric dipole moment of the molecule A, and $\mu=m_{M^+} m_A/(m_{M^+}+ m_A)$
is the reduced mass.

The back-dissociation microcanonical rate constant takes the simple form
\begin{eqnarray}
  K_B(E,J)={\sum_{J_A,J_B,\ell} P(E,J_A,\ell)\over h \rho_B(E,J)}
,\end{eqnarray}
where $J_A$ and $\ell$ are the angular momenta associated with molecule A and of M$^+$
with respect to A (with ${\bf J}={\bf J}_A+\boldmath{\ell}$), and
\begin{eqnarray}
  P(E,J_A,\ell)=\left\lbrace
  \begin{array}{c}
    0 \quad E < E_{J_A}\\
    0 \quad \ell > \sqrt{2\mu C}/\hbar\\
    1 \quad {\rm elsewhere}
.\end{array}
  \right.
\end{eqnarray}
The density of states of $(MA^+)^*$ depends on the binding energy
and the frequency of the normal modes, and contributes significantly to
slowing down the back-dissociation as it increases.

The radiative rate constant is given by \citep{Klippenstein1996}
\begin{eqnarray}\label{eq:Krad}
  K_E(E,J) =\sum_N P_N(E,J) A^J_N,
\end{eqnarray}
where $N={n_1, n_2, ... j, n_M}$ is a collective vibrational quantum number
specifying the vibrational states of the $n_M$ normal modes of  $(MA^+)^*$;
$ P_N(E,J)$ is the probability distribution of level $N$; and $A^J_N$ is the
radiative rate of level $N$ corresponding to the sum of the Einstein coefficients
over all possible lower states.
Due to the high density
of states, the sum over $N$ in Eq.~(\ref{eq:Krad}) is performed using a Monte Carlo sampling.
In the rigid rotor approximation and  summing  the H\"onl-London
factors {\citep{Whiting1974,Whiting1980}} over rotational transitions,
the vibrational emission rate   no longer
depends on $J$ and can  be
defined as
\begin{eqnarray}\label{eq:radiative-rate}
  A_N\approx\sum_{N'} {1\over 3\pi\epsilon_0 \hbar^4}\left( {E_N-E_{N'}\over c}\right)^3
  \vert M_{NN'}\vert^2,
\end{eqnarray}
where the  matrix elements in the normal mode approximation can be written
as
\begin{eqnarray}
  M_{NN'} 
  = \sum_{i=1}^{n_M} \sum_{\beta=x,y,z} {\partial d_\beta\over \partial Q_i}
  \sqrt{n_i\hbar \over \mu\omega_i}\quad \delta_{n_i,n_{i'-1}}\prod_{j\neq i} \delta_{n_j,n_j'}
,\end{eqnarray}
where only one vibrational quantum transition is considered because the
electric dipole component, $d_\alpha$, is approximated
around equilibrium, $Q_i$=0, as
\begin{eqnarray}\label{dipole-expansion}
  d_\alpha = d^0_\alpha + \sum_i Q_i \left.{\partial d_\alpha\over \partial Q_i}\right\vert_{Q_i=0} .
\end{eqnarray}

The parameters needed to obtain the
radiative association rate constant, $K_{RA}$, were calculated for MgC$_n$H$^+$ and MgC$_n$N$^+$.
The RCCSD(T)-F12
calculations were performed with the MOLPRO package \citep{Werner2020}
using the cvt-f12 basis \citep{Hill2010}. The geometries of all reactants and adducts
were optimized, and their normal modes were obtained. The
normal modes for MgC$_4$H$^+$($^1\Sigma$) and MgC$_3$N$^+$($^1\Sigma$)
are listed in Table~\ref{tab:MgC4H-normalmodes}.
The binding energies obtained, 3.78, 3.54, 3.58, and 3.44 eV (including zero-point energy)
for MgC$_3$N$^+$, MgC$_4$H$^+$, MgC$_5$N$^+$, and MgC$_6$H$^+$, respectively, do not vary
significantly among the four systems.
The dipole moments were calculated through calculations at different electric fields, and the derivatives
of Eq.~(\ref{dipole-expansion}) were obtained by a fit for each normal coordinate $Q_i$
(listed in Table~\ref{tab:MgC4H-normalmodes} for MgC$_4$H$^+$($^1\Sigma$) and MgC$_3$N$^+$($^1\Sigma$)).
Finally, the long-range interaction is characterized by the electric dipole moments,
2.89, 2.1, 3.39, and 5.54 debye for  C$_3$N($^2\Sigma$), C$_4$H($^2\Sigma$), C$_5$N($^2\Sigma$), and
C$_6$H($^2\Pi$), respectively.

\begin{table}[]
    \centering
    \caption{ MgC$_4$H$^+$ normal mode  frequencies (in cm$^{-1}$)  and first derivatives of the electric dipole, $d_\alpha$ ($\alpha$=x,y,z),  along each
    normal coordinate $Q_i$ (in atomic units).}
    \label{tab:MgC4H-normalmodes}
    \begin{tabular}{|c|ccc||ccc|}
      \hline
       & \multicolumn{3}{c}{MgC$_4$H$^+$} &  \multicolumn{3}{c}{MgC$_3$N$^+$}\\
      \hline
      i &       frequency  &  $\alpha$ & $\partial d_\alpha/\partial Q_i$ &   frequency  &  $\alpha$ & $\partial d_\alpha/\partial Q_i$  \\
      \hline
       1 &              76.13   & x &        -2.582e-3 &    84.45 & x & 3.10e-3\\
       2 &              76.19   & y &         2.582e-3 &    84.45 & y & 3.10e-3\\
       3 &             228.85   & x &        8.030e-4  &   245.74 & x & 4.64e-3\\
       4 &             228.86   & y &        8.030e-4  &   245.74 & y & 4.64e-3\\
       5 &             475.56   & z &       -1.020e-3  &   461.14 & z &  3.82e-3\\
       6 &             479.17   & x &        6.291e-4  &   489.14 & x & -7.49e-3\\
       7 &             479.17   & y &        6.291e-4  &   489.14 & y & -7.49e-3\\
       8 &             685.05   & x &         4.624e-3 &   972.58 & z & -2.37e-4\\
       9 &             685.06   & y &         4.624e-3 &  2110.48 & z &  2.43e-4\\
      10 &            1002.91   & z &         1.815e-3 &  2292.17 & z & -1.18e-3\\
      11 &            2059.33   & z &         3.672e-4 &- &- &- \\
      12 &            2212.11   & z &         4.900e-3 &- &- &- \\
      13 &            3436.51   & z &         7.036e-3 &- &- &- \\
     \hline
    \end{tabular}
  \end{table}

  The formation rate constants, $K_F(T)$, are very fast due to strong charge-electric dipole interaction
  (see left panels of Fig.~\ref{fig:rates}). The  differences are proportional to the particular
  dipole moments of the reactants, which vary at most by a factor of three, which is not appreciable
  at the scale of the figure.
  
  The emission rates, shown in the right panels of Fig.~\ref{fig:rates},
  show a progressive increase in the order  
  MgC$_3$N$^+$, MgC$_4$H$^+$, MgC$_5$N$^+$, and MgC$_6$H$^+$, essentially
  because the dipole moment of the adduct increases along this series,
  resulting in an increase of the $\partial d_\alpha/\partial Q_i$ derivatives.
  
  The back-dissociation rate, $K_B$, is dominated by the relatively high density of states, increasing
  with collision energy, as shown in Fig.~\ref{fig:rates}. The values of  $K_B(E)$ at low energies
  are very similar ($\approx$ 10$^6$ s$^{-1}$) except for MgC$_5$N$^+$ with
  $K_B(E)$ = 7\, 10$^{3}$ s$^{-1}$,
  which presents the lowest
  normal mode frequencies (of $\approx$ 54 cm${-1}$), and therefore it has the highest density of states.
  The case of MgC$_6$H$^+$ is special, showing $K_B(E)$
  very similar to the smaller systems of the series studied. 
  The C$_6$H reactant is a $^2\Pi$ state, while the  MgC$_6$H$^+$ adduct is  $^1\Sigma$,
  and there is a $\Pi$ state in the vicinity of its equilibrium geometry. 
  The $\Sigma-\Pi$ interaction introduces important perturbations in the frequencies of
  the normal modes.

  The ${K_E(T)/({K_E(T)+K_B(T)}})$ ratio decreases rapidly with collision energy, 
  as the number of open states   of the reactants increase, producing an increases in
  $K_B(T)$. The value at the lowest energy interval is particularly important to determine
  the radiative association rate constant, where $K_E(E)$ is closer to   $K_B(E)$, and
  the value of the ratio is particularly sensitive to the adduct density of states.

  {  
 The calculated total
 radiative association rate constants, $K_{RA}(T)$, are shown in the left panels of  Fig.~\ref{fig:rates}
 and listed in Table \ref{tab:rates} for different  temperatures. In general,  $K_{RA}(T)$ increases
 with the size of the system, being larger for MgC$_6$H$^+$ and MgC$_5$N$^+$ than for MgC$_4$H$^+$ and MgC$_3$N$^+$ according to  the trends explained above.
 For MgC$_n$N$^+$ the rates calculated here are within a factor of 5 of those obtained by \citet{Dunbar2002}
 for the related systems MgHC$_n$N$^+$ for the same $n$. At 10\,K, the rates
 are 4.5\, 10$^{-12}$ and 2.4\, 10$^{-11}$ cm$^3$ s$^{-1}$
 for MgC$_3$N$^+$ and MgHC$_3$N$^+$, while for $n$=5, the rates at 10\,K
 are 5.8\, 10$^{-9}$ and 2.4\, 10$^{-8}$ cm$^3$s$^{-1}$ for MgC$_5$N$^+$ and  MgHC$_5$N$^+$, respectively.
 These differences are partially due to the higher dipole moments of HC$_3$N and HC$_5$N (3.7 and 4.3 debye)
   compared to those of C$_3$N and C$_5$N (2.89 and 3.39 debye). The radiative association rate constant obtained for MgHC$_n$H$^+$ \citep{Dunbar2002}
 is considerably lower than that obtained here for MgC$_n$H$^+$ (a factor of $\approx$ 100 for MgHC$_4$H$^+$),
 simply because HC$_n$H has no permanent electric dipole, while C$_n$H does, favoring the formation of the adduct.

}

\begin{table}[]
    \centering
    \caption{Radiative association rate coefficients, $K_{RA}(T)$ in cm$^3$ s$^{-1}$,
      for the reactions of Mg$^+$ with C$_n$H and C$_n$N. These rates are fitted in the temperature range 10--300 K
    to the expression  $K_{RA}(T)$= A(cm$^3$ s$^{-1}$)\, ($T$/300 K)$^b$, and the parameters $A$ and $b$ are listed below. }
    \label{tab:rates}
    \begin{tabular}{ccccc}
\hline \hline
  $T$(K)  &      MgC$_4$H$^+$    &    MgC$_3$N$^+$   &    MgC$_6$H$^+$    &   MgC$_5$N$^+$  \\
\hline
    10  & 5.32\,10$^{-11}$ & 4.45\,10$^{-12}$ & 5.72\,10$^{-10}$ & 5.75\,10$^{-9}$  \\
    15  & 3.14\,10$^{-11}$ & 2.82\,10$^{-12}$ & 3.52\,10$^{-10}$ & 4.15\,10$^{-9}$  \\
    20  & 2.13\,10$^{-11}$ & 2.01\,10$^{-12}$ & 2.47\,10$^{-10}$ & 3.25\,10$^{-9}$  \\
    25  & 1.58\,10$^{-11}$ & 1.54\,10$^{-12}$ & 1.88\,10$^{-10}$ & 2.66\,10$^{-9}$  \\
    30  & 1.23\,10$^{-11}$ & 1.23\,10$^{-12}$ & 1.49\,10$^{-10}$ & 2.25\,10$^{-9}$  \\
    40  & 8.25\,10$^{-12}$ & 8.59\,10$^{-13}$ & 1.04\,10$^{-10}$ & 1.70\,10$^{-9}$  \\
    50  & 6.04\,10$^{-12}$ & 6.46\,10$^{-13}$ & 7.80\,10$^{-11}$ & 1.35\,10$^{-9}$  \\
   100  & 2.28\,10$^{-12}$ & 2.59\,10$^{-13}$ & 3.16\,10$^{-11}$ & 6.37\,10$^{-10}$ \\
   150  & 1.28\,10$^{-12}$ & 1.49\,10$^{-13}$ & 1.84\,10$^{-11}$ & 3.93\,10$^{-10}$ \\
   200  & 8.53\,10$^{-13}$ & 9.97\,10$^{-14}$ & 1.24\,10$^{-11}$ & 2.75\,10$^{-10}$ \\
   250  & 6.21\,10$^{-13}$ & 7.29\,10$^{-14}$ & 9.13\,10$^{-12}$ & 2.06\,10$^{-10}$ \\
   300  & 4.79\,10$^{-13}$ & 5.63\,10$^{-14}$ & 7.09\,10$^{-12}$ & 1.62\,10$^{-10}$ \\
& & & & \\
      \hline
   $A$    &   4.9\,10$^{-13}$ &  6.0\,10$^{-14}$ & 7.4\,10$^{-12}$ & 1.8\,10$^{-10}$ \\ 
   $b$    &   $-$1.39             &      $-$1.30         & $-$1.30.            &   $-$1.06          \\
\hline
    \end{tabular}
  \end{table}


\end{appendix}
\end{document}